\documentclass[journal, twoside]{IEEEtran}

\usepackage{enumerate}
\usepackage{multirow}
\usepackage{makecell}

\usepackage{cite}

\usepackage[pdftex]{graphicx}
\graphicspath{./Figures}
\usepackage{pgfplots}

\usepackage{amsmath}
\usepackage{amsthm}
\newtheorem{lemma}{Lemma}
\newtheorem{proposition}{Proposition}
\newtheorem{assertion}{Assertion}
\newtheorem{corollary}{Corollary}

\theoremstyle{definition}
\newtheorem{remark}{Remark}
\newtheorem{example}{Example}

\usepackage{amsfonts, amssymb, mathtools}
\usepackage{cases}

\usepackage{algorithm}
\usepackage{algorithmic}

\usepackage{array}
\newcolumntype{A}{>{$}r<{$}@{}>{${}}l<{$}}

\usepackage[caption=false,font=footnotesize]{subfig}

\usepackage{stfloats}

\usepackage{url}

\newcommand{\llbracket}{[\![}
\newcommand{\rrbracket}{]\!]}
\DeclareMathOperator{\im}{im}

\newcounter{storeeqcounter}
\newcounter{tempeqcounter}

\begin{document}
%
\title{Constructive Spherical Codes by Hopf Foliations}
%
%
\author{Henrique~K.~Miyamoto,~\IEEEmembership{Student Member,~IEEE,}
        Sueli~I.~R.~Costa,~\IEEEmembership{Member,~IEEE,}
        and~Henrique~N.~S\'a~Earp%
		\thanks{The work of H.~K.~Miyamoto was supported by S\~ao Paulo Research Foundation~(FAPESP) under grant 16/05126-0. The work of S.~I.~R.~Costa was supported by Brazilian National Council for Scientific and Technological Development~(CNPq) under grant 313326/2017-7 and by FAPESP under grant 13/25977-7. The work of H.~N.~S\'a~Earp was supported by CNPq under grant 307217/2017-5 and by FAPESP under grants 17/20007-0 and 18/21391-1. This paper was presented in part at the 2019 IEEE International Symposium on Information Theory~\cite{miyamoto}.}%
		\thanks{H.K.~Miyamoto was with the School of Electrical and Computer Engineering (FEEC), University of Campinas (Unicamp), Campinas, SP 13083-852 Brazil. He is now with the Institute of Mathematics, Statistics and Scientific Computing (IMECC), University of Campinas (Unicamp), Campinas, SP 13083-859 Brazil (e-mail: miyamotohk@gmail.com).}%
		\thanks{S.~I.~R.~Costa and H.~N.~S\'a~Earp are with the Institute of Mathematics, Statistics and Scientific Computing (IMECC), University of Campinas (Unicamp), Campinas, SP 13083-859 Brazil (e-mail: sueli@unicamp.br; henrique.saearp@ime.unicamp.br).}%
		\thanks{Copyright~\textcopyright~2021 IEEE. Personal use of this material is permitted.  However, permission to use this material for any other purposes must be obtained from the IEEE by sending a request to pubs-permissions@ieee.org.}%
}
%

\IEEEpubid{ }
\maketitle

\begin{abstract}
	We present a new systematic approach to constructing spherical codes in dimensions $2^k$, based on Hopf foliations. Using the fact that a sphere $S^{2n-1}$ is foliated by manifolds $S_{\cos\eta}^{n-1} \times S_{\sin\eta}^{n-1},\ \eta\in\left[0,\pi/2\right]$, we distribute points in dimension $2^k$ via a recursive algorithm from a basic construction in $\mathbb{R}^4$. Our procedure outperforms some current constructive methods in several small-distance regimes and constitutes a compromise between achieving a large number of codewords for a minimum given distance and effective constructiveness with low encoding computational cost. Bounds for the asymptotic density are derived and compared with other constructions. The encoding process has storage complexity $O(n)$ and time complexity $O(n \log n)$. We also propose a sub-optimal decoding procedure, which does not require storing the codebook and has time complexity $O(n \log n)$.
\end{abstract}

\begin{IEEEkeywords}
	Asymptotic density, encoding and decoding complexity, Hopf foliation, spherical codes.
\end{IEEEkeywords}

%
\IEEEpeerreviewmaketitle

\section{Introduction}
%
\IEEEPARstart{A}{spherical} code $\mathcal{C}(M,n,d) \coloneqq \{x_1,x_2,\dots,x_M\} \subset S^{n-1}$ is a set of $M$ points on the unit Euclidean sphere in $\mathbb{R}^n$ with minimum Euclidean distance at least $d$, cf.~\cite{ericson}. Problems with spherical codes involve finding optimal distributions of points relative to some parameter of interest, and they lend themselves to several applications. From a practical point of view, it is also desirable that a code exhibits algebraic constructions or geometric regularities, which can provide lower complexity in the encoding and decoding processes. The spherical packing problem in spherical code design can be considered in the following presentation: given a minimum Euclidean distance $d>0$, to find the largest possible number $M$ of points on $S^{n-1}$ with all mutual distances at least $d$. The solution is trivial for $n=2$, namely a regular polygon, but few optimal solutions are known for higher dimensions. Special codes and some best known codes for a given distance in selected dimensions are presented in \cite{ericson} and \cite{sloane-table}.

Among the most well-known constructive spherical codes, we highlight the so-called \emph{apple-peeling}~\cite{elgamal}, \emph{wrapped}~\cite{hamkins1} and \emph{laminated}~\cite{hamkins2} methods, the last two being asymptotically dense. The \emph{torus layers spherical codes}~(TLSC)~\cite{torezzan}, while not asymptotically dense, have a more homogeneous structure, in the sense that points on the same leaf are indistinguishable with respect to distance profile, and have been shown to compare favorably with other codes for non-asymptotic minimum distances. This method foliates the sphere $S^{2n-1}$ by flat manifolds $S^1\times\dots\times S^1$ and distributes points using good packing density lattices in the half-dimension~\cite{conway, costa}. Other recent contributions to this topic include codes obtained by partitioning the sphere into regions of equal area~\cite{leopardi}, bounds for constructible codes near the Shannon bound~\cite{sole}, commutative group codes~\cite{siqueira, alves} and cyclic group codes~\cite{taylor}. One main challenge for the application of spherical codes is the effective constructiveness for a large range  of distances at a reasonable computational cost, which we propose to address in this work.

Classical applications of spherical codes  in communications include channel coding, as a generalization of PSK modulation, and source coding, using shape-gain vector quantizers~\cite{hamkins4, boaventura}. The problem of optimal constellation design for signalling in non-coherent communications can be formulated as a sphere packing on the Grassmannian manifold of lines~\cite{zheng}, which, in turn, is associated to an antipodal spherical code~\cite{conway2}. A recent example of such approach can be found in~\cite{ngo}. Furthermore, spherical codes have been used in schemes to improve power efficiency of communication systems in the context of MIMO communications~\cite{sedaghat, rachinger}.

In the context of coherent optical communications, four-dimensional modulations have been considered in order to exploit the physical nature of the electromagnetic field. In~\cite{agrell, karlsoon1, rodrigues}, the performance of four-dimensional modulations is studied and spherical codes are also considered. In~\cite{karlsson2}, the authors observe that, at low spectral efficiencies, in dimensions two and four, spherical codes have optimal or close to optimal performance. The performance of modulations in dimensions 8 and 16 has also been addressed in~\cite{reimer,rademacher}.

\IEEEpubidadjcol

We propose a construction of spherical codes inspired by the TLSC method and the Hopf fibration, which gives a somewhat `natural' foliation of $S^3$, $S^7$ and $S^{15}$, and which also appears in problems in physics and communications~\cite{rodrigues, urbantke, mosseri}. Our procedure exploits Hopf foliations in dimensions $2^k$ to construct a family of \emph{spherical codes by Hopf foliations}~(SCHF), by means of a recursive algorithm for any given minimum distance $d \in \left]0,2\right]$. The initial step is a flat model in $\mathbb{R}^{4}$, for which this construction is equivalent to TLSC via special lattices in dimension two. For higher dimensions, the construction is qualitatively different and, besides defining a much simpler algorithm, for certain minimum distances, it outperforms known TLSC implementations in terms of code cardinality. Although we focus on codes in dimensions $2^k$ with basic dimension 4, the procedure presented here can be applied to any even dimension $2n$, if provided with a family of spherical codes in dimension $n$. The performance analysis of the proposed codes includes the comparison with other known constructions, determining their asymptotic density, and computing the complexity of the encoding and decoding processes.

This paper is organized as follows: Section~\ref{sec:hopf} is an introduction to Hopf foliations. Section~\ref{sec:construction} introduces the SCHF, and derives some of their properties and the recursive construction procedure. In Section~\ref{sec:performance}, we present numerical results for constructions in dimensions 4, 8, 16, 32 and 64. In Section~\ref{sec:density}, we derive asymptotic density bounds for our family of codes, which can be closely approached in the simulations in Section~\ref{sec:performance}. Section~\ref{sec:complexity} discusses the encoding complexity, showing that this construction has storage complexity $O(n)$ and time complexity $O(n\log n)$. In Section~\ref{sec:decoding}, we provide a suboptimal decoding algorithm with time complexity $O(n \log n)$ and storage complexity $O(1)$, which avoids the high-complexity of the ML decoder, while keeping reasonable decoding performance in terms of error rate. Finally, in Section~\ref{sec:conclusion}, we draw some conclusions and perspectives for subsequent work.

\section{Hopf Fibration and Sphere Foliations} \label{sec:hopf}

\begin{figure}[!t]
	\centering
	\includegraphics[height=7cm]{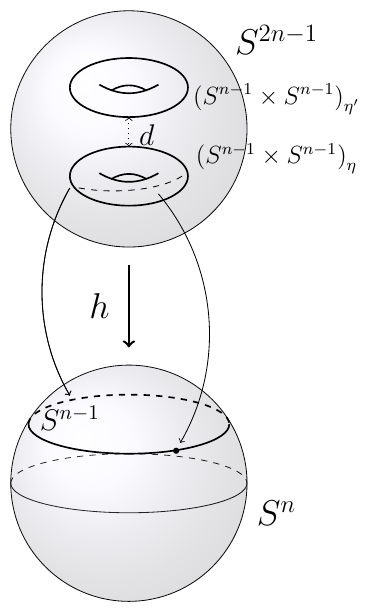}
	\caption{The generalized Hopf map.}
	\label{fig:hopf-map}
\end{figure}

We denote the Euclidean sphere at the origin of $\mathbb{R}^n$, with radius $r$, by
\begin{displaymath}
	S_r^{n-1} \coloneqq \{ \mathbf{x} \in \mathbb{R}^n : \|\mathbf{x}\| = r \},
\end{displaymath}
and the unit sphere simply by $S^{n-1} \coloneqq S_{1}^{n-1}$. In real dimensions $n\in\{1,2,4,8\}$, let $\mathbb{A}\cong\mathbb{R}^n$ be the corresponding normed division algebra: respectively, the real numbers~$\mathbb{R}$, the complex numbers~$\mathbb{C}$, the quaternions~$\mathbb{H}$ or the octonions~$\mathbb{O}$, cf.~\cite{baez}. Identifying $\mathbb{R}^{2n} \cong \mathbb{A}^2$ by $(x_1,\dots,x_{n};x_{n+1},\dots,x_{2n}) \leftrightarrow (z_0,z_1)$ and $\mathbb{R}^{n+1} \cong \mathbb{A}\times\mathbb{R}$ by $(x_1, \dots, x_n;x_{n+1}) \leftrightarrow (z;x_{n+1})$, the unit $(2n-1)$- and $n$-spheres can be described respectively by
\begin{equation} \label{eq: Sphere S^(2n-1)}
	S^{2n-1} = \{ \left(z_0,z_1\right) \subset\mathbb{A}^2 : |z_0|^2+|z_1|^2=1 \}
\end{equation}
and
\begin{displaymath}
	S^n = \{ \left(z;x_{n+1}\right) \subset \mathbb{A}\times\mathbb{R} : |z|^2+x_{n+1}^2=1 \}.
\end{displaymath}
In this description, for $n \in \{1,2,4,8\}$, the \emph{Hopf fibration}~\cite{lyons, nakahara} is the (submersion) map
\begin{align}  \label{eq: Hopf fibration}
	\begin{split}
	h: S^{2n-1} &\to S^{n}\\
	\left( z_0,z_1 \right) &\mapsto \left( 2z_0\overline{z_1},|z_0|^2-|z_1|^2 \right)
	\end{split}
\end{align}
in which $z_0,\ z_1\in\mathbb{A}$ (see Fig.~\ref{fig:hopf-map}).

Since $|z_0|^2+|z_1|^2=1$, there is a unique $\eta \in \left[0,\pi/2\right]$ such that  $|z_0|=\cos\eta$ and $|z_1|=\sin\eta$. Each value of $\eta$ determines a height $x_{n+1} = |z_0|^2 - |z_1|^2 = \cos 2\eta$ in the image $S^n$, cutting out a $(n-1)$-sphere $S_{\sin 2 \eta}^{n-1} \subset S^n$ to which we refer as a \emph{parallel slice}. Furthermore, the preimage $h^{-1}(P)$ of each point $P\in S^n$ under the Hopf fibration (\ref{eq: Hopf fibration}) is a great sphere $S^{n-1} \subset S^{2n-1}$, called the \emph{fiber of $h$ over $P$}.

Varying $P$ over such a parallel slice spans a preimage  in the total space $S^{2n-1}$ comprising the union of the corresponding fibers, and it can thus be described as the product $S_{\cos\eta}^{n-1}\times S_{\sin\eta}^{n-1} \subset S^{2n-1}$. Hence, by considering all parallel slices in the base sphere $S^n$, we characterize $S^{2n-1}$ as a disjoint union of product manifolds
\begin{displaymath}
	T^{2n-2}_{\eta} \coloneqq \left(S^{n-1}\times S^{n-1}\right)_{\eta} \coloneqq S_{\cos\eta}^{n-1}\times S_{\sin\eta}^{n-1}.
\end{displaymath}
This decomposition is an instance of a \emph{foliation}, the \emph{leaves} of which are the generalized tori $T^{2n-2}_{\eta}$; in other words, the sphere $S^{2n-1}$ is said to be \emph{foliated} by tori $T^{2n-2}_{\eta}$. So we will incorporate this vocabulary from differential topology, but we will not invoke any substantial results from that theory in this paper.

As it turns out, this structure, which we call \textit{Hopf foliation}, is not restricted to the cases $n\in\{1,2,4,8\}$, indeed it extends to any $n\in\mathbb{N}^*$, regardless of the existence of an associated normed division algebra in that dimension:

\begin{assertion} \label{assertion:general}
	For every $n\in\mathbb{N}^*$, the sphere $S^{2n-1}\subset\mathbb{R}^{2n}$ is foliated by manifolds $T^{2n-2}_{\eta} = \left(S^{n-1}\times S^{n-1}\right)_{\eta}$.
\end{assertion}

Explicitly, write $\mathbf{x} = \left(x_1,\dots,x_{2n}\right) \in  S^{2n-1}\subset\mathbb{R}^{2n}$ as
\begin{equation}
	\mathbf{x} = \left( \alpha\frac{(x_1,\dots,x_n)}{\alpha};\,\beta\frac{(x_{n+1},\dots,x_{2n})}{\beta} \right)
\end{equation}
for $\alpha \coloneqq \|(x_1,\dots,x_n)\|$, $\beta \coloneqq \|(x_{n+1},\dots,x_{2n})\|$ and $\alpha,\,\beta\neq0$. For $\alpha=0$ or $\beta=0$, we have degenerate manifolds $\mathbf{0} \times S_{\sin\eta}^{n-1}$ or $S_{\cos\eta}^{n-1} \times \mathbf{0}$. Since for any $\mathbf{x}\in S^{2n-1}$ we have $\alpha^2+\beta^2=1$,  there is a unique $\eta\in\left[0,\pi/2\right]$ such that $\alpha=\cos\eta$ and $\beta=\sin\eta$, so
\begin{displaymath}
	\mathbf{x} = \left( \cos\eta\,\mathbf{v}_1;\,\sin\eta\,\mathbf{v}_2 \right),\quad \mathbf{v_1},\mathbf{v_2}\in S^{n-1}.
\end{displaymath}
This describes the foliation of the unit sphere $S^{2n-1}\subset\mathbb{R}^{2n}$ by products of spheres $S^{n-1}_{\cos\eta}\times S^{n-1}_{\sin\eta}$ of radii $\cos\eta$ and $\sin\eta$.

In particular, the Hopf fibration in dimension 4 $(n=2)$ gives a foliation of $S^3$ by two-dimensional flat tori $T_\eta^2 = S_{\cos\eta}^1\times S_{\sin\eta}^1$:
\begin{align} \label{eq: iota S^3 por T^2}
	\begin{split}
	\iota: \left[0,\frac{\pi}{2}\right] \times {\left[0,2\pi\right[}^{2}\; &\to\; S^3 \\
	(\eta;\xi_1,\xi_2)\; &\mapsto\; \mkern-50mu
	\underbrace{\left(e^{\boldsymbol{i}\xi_1}\cos\eta,\,e^{\boldsymbol{i}\xi_2}\sin\eta\right).}_{\cong \left(\cos\eta\, (\cos\xi_1, \sin\xi_1);\; \sin\eta\, (\cos\xi_2, \sin\xi_2)\right)}
	\end{split}	
\end{align}

For each angle $\eta\in\left[0,\pi/2\right]$, the induced map 
\begin{equation} \label{eq: iota-eta}
	\iota_\eta:[0,2\pi[^2 \to S^3
\end{equation}
spans the $2$-torus 
\begin{equation} \label{eq: T-eta}
	T_\eta \coloneqq T_{\eta}^2 = S_{\cos\eta}^1\times S_{\sin\eta}^1=\im \iota_\eta\subset S^3
\end{equation}
of Euclidean radii $\cos\eta$ and $\sin\eta$. When $\eta\in\{0,\pi/2\}$, the parametrization describes circles, which are degenerate tori. Taking $\eta\notin\{0,\pi/2\}$ and $\mathbf{c}=(c_1,c_2) \coloneqq (\cos\eta,\sin\eta)$, the image  $\iota_\eta(u/\cos\eta,v/\sin\eta)=\Phi_{\mathbf{c}}(u,v)$ coincides with the flat tori map defined in~\cite{torezzan}. Moreover, $\iota_\eta({\left[0,2\pi\right[}^{2})=\Phi_\mathbf{c}(\left[0,2\pi c_1\right[ \times \left[0, 2\pi c_2\right[)$ and $\Phi_{\mathbf{c}}$ is a local isometry, which maps the rectangle into the flat torus in $\mathbb{R}^4$ by gluing its parallel boundary segments.

\section{Construction of Spherical Codes} \label{sec:construction}

Our construction of spherical codes, inspired by the Hopf fibration, uses the foliations of Assertion~\ref{assertion:general} to algorithmically distribute points on spheres $S^{2n-1} \subset \mathbb{R}^{2n}$, given a minimum mutual Euclidean distance $d \in \left]0,2\right]$. Each part of the code constructed on a Cartesian product $\left(S^{n-1}\times S^{n-1}\right)_{\eta}$ corresponds to a \textit{direct sum} \cite[Section 1.7]{ericson} of codes on each copy of $S^{n-1}$. We construct each code $\mathcal{C}(M,n,d)$ as the union of several such products.

\subsection{Choosing the Leaves}

The next result, obtained by straightforward calculation, is used to choose the layers of leaves $\left(S^{n-1}\times S^{n-1}\right)_{\eta}$, all along our recursive procedure. We remark that, restricted to $n=2$, this proposition is the same as \cite[Proposition 1]{torezzan}. We denote henceforth the integer points of an interval $\left[a,b\right]$ by  $\llbracket a, b \rrbracket \coloneqq \left[a, b\right] \cap \mathbb{Z}$.

\begin{proposition} \label{prop:dist-tori}
	The minimum distance between two leaves	$T^{2n-2}_{\eta} = \left(S^{n-1}\times S^{n-1}\right)_\eta$ and $T^{2n-2}_{\eta'} = \left(S^{n-1}\times S^{n-1}\right)_{\eta'}$ is
	\begin{equation} \label{eq: min distance d}
		d\left(T^{2n-2}_{\eta},T^{2n-2}_{\eta'}\right) = 2\sin\left(\frac{\eta-\eta'}{2}\right),
	\end{equation}
	which coincides with the Euclidean distance between two points of angles $\eta$ and $\eta'$ on the first quadrant of $S^1$.
\end{proposition}

\begin{figure}[t]
	\centering
	\includegraphics[width=7cm]{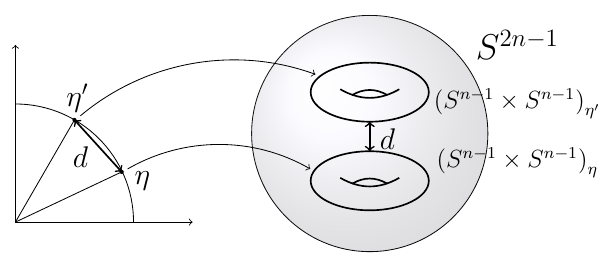}
	\caption{The distance between leaves $T^{2n-2}_\eta=\left(S^{n-1}\times S^{n-1}\right)_{\eta}$ and $T^{2n-2}_{\eta'} = \left(S^{n-1} \times S^{n-1}\right)_{\eta'}$ in $\mathbb{R}^{2n}$, viewed as a chordal distance between points determined by the angles $\eta$ and $\eta'$ in $S^1$.}
	\label{fig: S1}
\end{figure}

\begin{IEEEproof}
	Adopting the notation $\mathbf{v}_1 \coloneqq (v_1,\dots,v_n)$ and $\mathbf{v}_2 \coloneqq (v_{n+1},\dots,v_{2n})$, take two points
	\begin{IEEEeqnarray*}{rCl}
		\mathbf{x} &=& \left( \cos\eta\,\mathbf{v}_1;\, \sin\eta\,\mathbf{v}_2 \right) \in \left(S^{n-1}\times S^{n-1}\right)_\eta
		\\
		\noalign{\noindent and \vspace{2\jot}}
		\mathbf{x}' &=& \left( \cos\eta'\,\mathbf{v}_1';\, \sin\eta'\,\mathbf{v}_2' \right) \in \left(S^{n-1}\times S^{n-1}\right)_{\eta'},
	\end{IEEEeqnarray*}
	with $\|\mathbf{v}_i\| = \|\mathbf{v}_i'\| = 1$, for $i\in\{1,2\}$. For the squared Euclidean distance $d^2(\mathbf{x},\mathbf{x'})$ we have	
	\begin{align*}
		d^2(\mathbf{x},\mathbf{x'}) &= \| \mathbf{x} - \mathbf{x'} \|^2 = \|\mathbf{x}\|^2 + \|\mathbf{x}'\|^2 - 2 \left\langle \mathbf{x}, \mathbf{x}' \right\rangle \\
		&= 2 - 2 \left(\cos\eta\cos\eta'\langle \mathbf{v}_1, \mathbf{v}_1' \rangle + \sin\eta\sin\eta'\langle \mathbf{v}_2, \mathbf{v}_2' \rangle\right)\\
		&\ge 2 -2 \left(\cos\eta\cos\eta'+\sin\eta\sin\eta'\right)\\
		&=2 \left[1-\cos(\eta-\eta')\right]\\
		&=2 \left[1 -\left(1 - 2\sin^2 \left(\frac{\eta-\eta'}{2}\right) \right) \right]\\
		&=4\sin^2 \left(\frac{\eta-\eta'}{2}\right),
	\end{align*}
	and equality holds if, and only if, $\mathbf{v}_1 = \mathbf{v}_1'$ and $\mathbf{v}_2 = \mathbf{v}_2'$. Therefore the minimum distance between the sets $\left(S^{n-1}\times S^{n-1}\right)_\eta$ and $\left(S^{n-1}\times S^{n-1}\right)_{\eta'}$ is $d\left(T^{2n-2}_\eta,T^{2n-2}_{\eta'}\right) = 2\sin\left(\frac{\eta-\eta'}{2}\right)$, which is the chordal distance between points determined by angles $\eta$ and $\eta'$ on the first quadrant of the circle $S^1 \subset \mathbb{R}^2$ (see Fig.~\ref{fig: S1}).
\end{IEEEproof}

\begin{corollary} \label{cor: t}
	In the context of Proposition~\ref{prop:dist-tori}:
	\begin{enumerate}[a)]
		\item The minimum angular interval between $\eta$ and $\eta'$ respecting the minimum distance $d$ is
		\begin{equation} \label{eq: Deta}
			\Delta\eta \coloneqq |\eta-\eta'| = 2\arcsin(d/2).
		\end{equation}
		
		\item The maximum number of leaves separated by distance $d$ is $t(d)+1$, with
		\begin{equation} \label{eq: t}
			t(d)= \left\lfloor\frac{\pi}{4\arcsin(d/2)}\right\rfloor.
		\end{equation}
		
		\item \label{eq: etas}
		We may choose the leaves $S_{\cos\eta}^{n-1}\times S_{\sin\eta}^{n-1}$ separated by at least $d$, considering
		\begin{enumerate}[i)]
			\item $\eta=\eta_0 + k \Delta\eta$, for $k\in\llbracket0,t(d)\rrbracket$ and $0\le\eta_0\le(\pi/2-t(d)\Delta\eta)/2$, or
			\item $\eta=\pi/4 \pm k \Delta\eta$, for $k\in\llbracket0,\lfloor t(d)/2 \rfloor \rrbracket$.
		\end{enumerate}
		In the latter case, leaves are symmetrically chosen around $\eta=\pi/4$, the leaf of greatest `area'.
	\end{enumerate} 
\end{corollary}

Once the leaves have been chosen with a guaranteed minimum mutual distance $d$, we proceed to construct a spherical code in $S^{2n-1}$, by considering codes on each leaf $T^{2n-2}_{\eta}$ with the desired minimum distance. We illustrate this idea with an example, to show that there are several ways of choosing these leaves.

\begin{example}
	For minimum distance $d=1$, we have $\Delta\eta = \pi/3$ and $t(1) = 1$. We can choose different sets of leaves, for instance, $\eta = \frac{\pi}{4}$, $\eta \in \{0, \frac{\pi}{3}\}$ or $\eta \in \{\frac{\pi}{12},\frac{5\pi}{12}\}$. In dimension $n=4$, we can construct one of the following codes:
	\begin{enumerate}
		\item \label{ex1-item1}
		Case $\eta = \frac{\pi}{4}$ (only one leaf). Consider the code in $S^{1}_{1/{\sqrt{2}}} \times S^{1}_{1/{\sqrt{2}}}$ as the product code $\mathcal{C} = \mathcal{C}_{\text{bi}} \times \mathcal{C}_{\text{bi}}$, where $\mathcal{C}_{\text{bi}}$ is the biorthogonal code in $S^{1}_{{1}/{\sqrt{2}}}$, given as the set of all permutations of $(\pm\sqrt{1/2}, 0)$, which has minimum distance $1$ and 16 codewords.
		
		\item \label{ex1-item2}
		Case $\eta\in\{0,\frac{\pi}{3}\}$ (two leaves). For $\eta=0$, consider the code $\mathcal{C}_1 = \mathcal{C}_{\text{hex}} \times \{(0,0)\}$, where $\mathcal{C}_{\text{hex}}$ is the hexagonal code in $S^1_1$. For $\eta=\frac{\pi}{3}$, consider $\mathcal{C}_2=\mathcal{C}_{\text{anti}} \times \mathcal{C}_{\text{pen}}$, where $\mathcal{C}_{\text{pen}}$ is the pentagon in $S^1_{{\sqrt{3}}/{2}}$ and $\mathcal{C}_{\text{anti}}$ is the set of two antipodal points in $S^1_{{1}/{2}}$. The final code $\mathcal{C} = \mathcal{C}_1 \cup \mathcal{C}_2$ is a spherical code with 16 codewords.
		
		\item \label{ex1-item3}
		For $\eta\in\{ \frac{\pi}{12}, \frac{5\pi}{12} \}$ (two symmetrical leaves), consider the codes $\mathcal{C}_1 = \mathcal{C}_{\text{pen}}\times\mathcal{C}_{\text{one}}$ and $\mathcal{C}_2 = \mathcal{C}_{\text{one}}\times\mathcal{C}_{\text{pen}}$, where $\mathcal{C}_{\text{one}}$ is a single-point code in $S^1_{\sin(\pi/12)} = S^1_{\cos(5\pi/12)}$ and $\mathcal{C}_{\text{pen}}$ is the pentagonal code in $S^1_{\cos(\pi/12)} = S^1_{\sin(5\pi/12)}$. Each $\mathcal{C}_1$, $\mathcal{C}_2$ has 5 points, and the final code $\mathcal{C} = \mathcal{C}_1 \cup \mathcal{C}_2$ has 10 points, which is less than the 12 points obtained by taking $\eta\in\{0,\frac{\pi}{2}\}$ (degenerate tori).
		
		\item \label{ex1-item4}
		Note that Proposition~\ref{prop:dist-tori} provides a sufficient condition for the minimum distance. But in this case we can see that, besides the codewords of item~\ref{ex1-item1}), we can also consider points in the degenerate tori defined by $\eta\in\{0,\frac{\pi}{2}\}$, i.e., the codewords $(0,0,\pm\sqrt{1/2},\pm\sqrt{1/2})$ and $(\pm\sqrt{1/2},\pm\sqrt{1/2},0,0)$, and still have minimum distance $1$ between the 24 codewords. This spherical code is the best known for $d=1$ in $\mathbb{R}^4$ \cite{ericson} and a similar code for this distance with $4\binom{n}{2}$ codewords can be obtained in $\mathbb{R}^{n}$.
	\end{enumerate}
\end{example}

In the algorithm we will formulate shortly, we set a procedure based on Proposition~\ref{prop:dist-tori}, considering different choices for the leaves, such as the ones in items \ref{ex1-item1}), \ref{ex1-item2}) and \ref{ex1-item3}), and a few special codes such as the one in item~\ref{ex1-item4}). These examples illustrate the fact that the general construction by leaves is complex, as it requires choosing good codes in the half-dimension. This motivates us to propose a recursive procedure for dimensions $2^k$, using dimension $n=4$ as the basic case.

\subsection{Basic Case: Spherical Codes in $\mathbb{R}^4$}

Given a minimum distance $d \in \left]0,2\right]$, our procedure is based on a two-step process:
\begin{enumerate}
	\item Choose a set of parameters $H = \{\eta_1, \dots, \eta_p\} \subset [0,\pi/2]$, generating a family of tori $\{T_\eta=S_{\cos\eta}^1\times S_{\sin\eta}^1 : \eta \in H\}$ mutually distant by at least~$d$, as sets in $\mathbb{R}^4$, cf. Corollary~\ref{cor: t}--\ref{eq: etas}).
	
	\item On each torus~$T_\eta$, distribute points with minimal mutual distance~$d$, following three steps:
	\begin{enumerate}
		\item 
		choose $n$ \emph{internal circles}, i.e., the images $\iota_\eta(\xi_1,\xi_2)$, for $\xi_1\in\left[0,2\pi\right[$ and fixed $\xi_2$, mutually distant by at least $d$ and separated by $\Delta\xi_2$;
		\item 
		on each such circle, distribute $m$ equidistant points, separated by $\Delta\xi_1$;
		\item
		shift the distributions of consecutive internal circles by $\Delta\xi_1/2$, so as to bring those circles closer and improve the point density.
	\end{enumerate}
\end{enumerate}

\begin{figure}[!t]
	\centering
	\subfloat[]{\includegraphics[height=4.5cm]{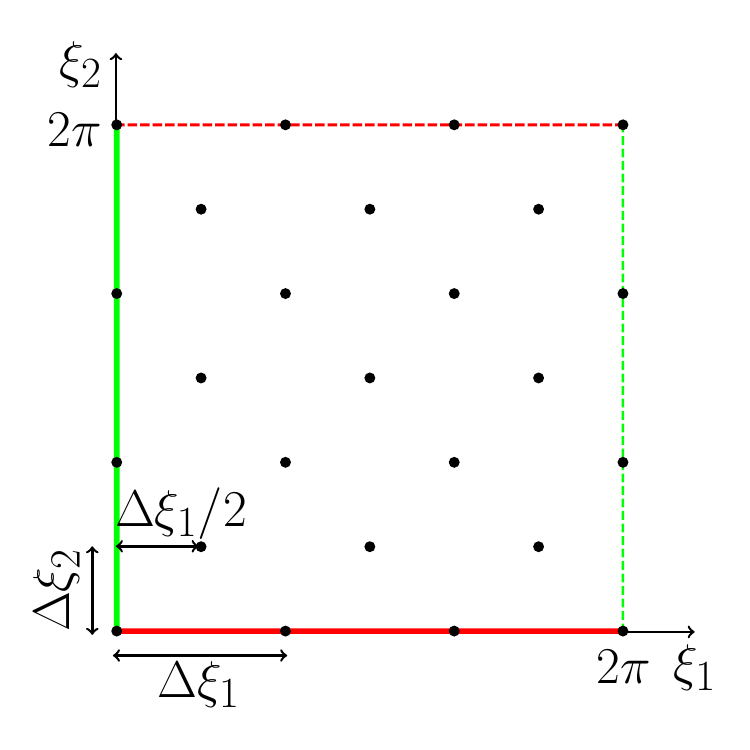}}
	\label{fig: lattice-a}
	\hfill
	\subfloat[]{\includegraphics[height=4.5cm]{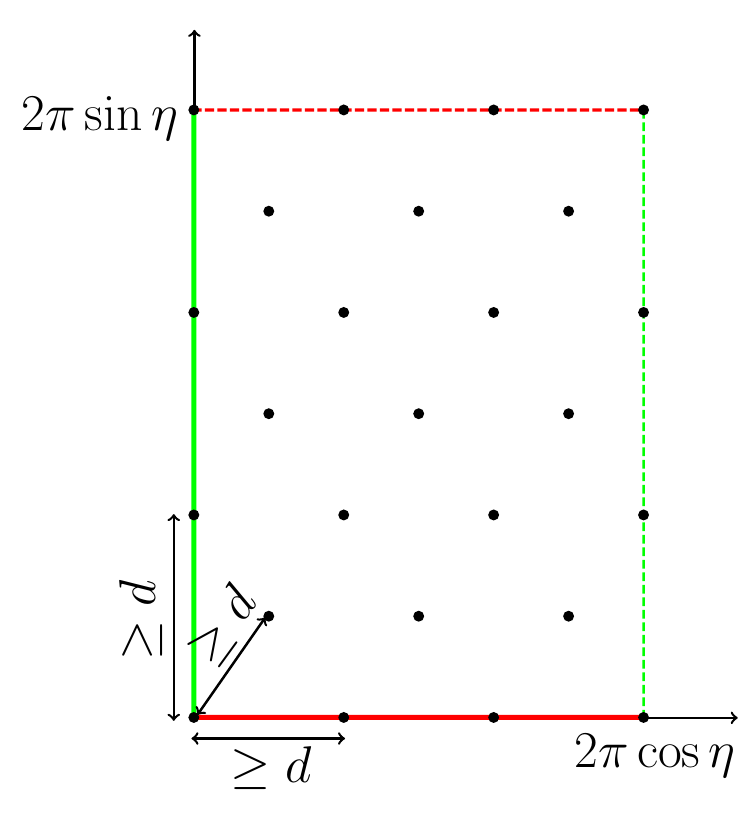}}
	\label{fig: lattice-b}
	\caption{Distribution of points on a torus $T_{\eta}$, seen as the preimage by $\iota_{\eta}$ (a) and $\Phi_{\mathbf{c}}$ (b), where $\mathbf{c}=(\cos\eta, \sin\eta)$.	In this case, $d=0.9$ and $\eta=2\arcsin(d/2)$, so that $m=3$ and $n=6$. Moreover, $\Delta\xi_1 = 2\pi/m = 2\pi/3$ and $\Delta\xi_2 = 2\pi/n = \pi/3$. Note the displacement of $\Delta\xi_1/2$ between consecutive internal circles.}
	\label{fig: lattice}
\end{figure}

An illustration of these parameters is given in Fig.~\ref{fig: lattice}. The next result provides a way to determine the number~$n$ of internal circles and the number~$m$ of points on each such circle, within each~$T_\eta$.

\begin{proposition} \label{prop:T_eta}
	On each torus $T_\eta$, defined as in \eqref{eq: T-eta}, for $\eta \in \left[0,\pi/2\right]$, we have:
	\begin{enumerate}[a)]
		\item \label{prop:m}
		The maximum number $m=m(d,\eta)$ of points that can be distributed on a internal circle, respecting the minimum mutual distance $d$, is
		\begin{equation}
		\label{eq: MNP on a circle m}
		m(d,\eta)=
		\begin{cases}
		\left\lfloor {\dfrac{\pi}{\arcsin\big( d/(2\cos\eta) \big)}} \right\rfloor,& \mbox{if}\quad d\le2\cos\eta,\\
		1, &\mbox{otherwise}.
		\end{cases}
		\end{equation}
		
		\item \label{prop:n}
		The maximum number $n=n(d,\eta)$ of internal circles that can be distributed on $T_\eta$, with shifting angle $\pi/m$, $m=m(d,\eta)$, such that the mutual distance among their points is at least $d$, is
		\begin{equation}
		\label{eq: MN circles n}
		n(d,\eta)=\max\{\tilde{n},1\},
		\end{equation}
		with $\tilde{n}=2\lfloor\min\{n_1,n_2\}/2\rfloor$ and
		\begin{align}        
		&n_1=\left\lfloor \dfrac{\pi}{\arcsin\left[\big((d^2/4)\csc^2\eta-\cot^2\eta\sin^2(\pi/2m)\big)^{\frac{1}{2}}\right]} \right\rfloor\\
		&n_2=
		\begin{cases}
		\left\lfloor \frac{2\pi}{\arcsin\big(d/ (2\sin\eta) \big)} \right\rfloor,& \mbox{if}\quad d\le2\sin\eta,\\
		1, &\mbox{otherwise}.
		\end{cases}
		\end{align}
	\end{enumerate}
\end{proposition}

\begin{IEEEproof}
	The calculations are straightforward using that the squared distance between the image by $\iota_\eta$ as in \eqref{eq: iota-eta} of two points determined by angles $(\xi_1,\xi_2)$ and $(\xi_1',\xi_2')$ is
	\begin{align} \label{eq: d-iota-eta-general}
		\hspace{2em}&\hspace{-2em}d^2\big(\iota_\eta(\xi_1,\xi_2),\iota_\eta(\xi_1',\xi_2')\big) \nonumber\\
		&= |e^{\boldsymbol{i}\xi_1}\cos\eta-e^{\boldsymbol{i}\xi_1'}\cos\eta|^2 + |e^{\boldsymbol{i}\xi_2}\sin\eta-e^{\boldsymbol{i}\xi_2'}\sin\eta|^2.
	\end{align}
	
	In particular, when the points are in the same internal circle determined by $\xi_2$ and are displaced by $\Delta\xi_1$, i.e., $\xi_1'=\xi_1 + \Delta\xi_1$ and $\xi_2'=\xi_2$, we have
	\begin{equation} \label{eq: d on same circle}
		d\big(\iota_\eta(\xi_1,\xi_2),\iota_\eta(\xi_1+\Delta\xi_1,\xi_2)\big) = 2\cos\eta\sin(\Delta\xi_1/2).
	\end{equation}
	Note that $m = m(d,\eta)$ is obtained by ensuring minimum distance between points in the same internal circle. Using that $m = \lfloor2\pi/\Delta\xi_1\rfloor$, where $\Delta\xi_1=2\arcsin\big(d/(2\cos\eta)\big)$ is obtained by inverting~\eqref{eq: d on same circle} and setting the distance to $d$, yields~\eqref{eq: MNP on a circle m}.

	Now, for $n=n(d,\eta)$, we have to ensure minimum distance between points both in shifted and aligned internal circles (see Fig.~\ref{fig: lattice}). First, we compute the distance between two points in the image by $\iota_\eta$, one in each of two consecutive internal circles, with point distributions shifted by $\Delta\xi_1/2=\pi/m$, with $m=m(d,\eta)$. Hence, setting $\xi_1' = \xi_1 + \Delta\xi_1/2$ and $\xi_2' = \xi_2 + \Delta\xi_2$ in~\eqref{eq: d-iota-eta-general} gives
	\begin{align}\label{eq: d displaced circles}
		\hspace{3em}&\hspace{-3em}d\big(\iota_\eta(\xi_1,\xi_2),\,\iota_\eta(\xi_1+\Delta\xi_1/2,\xi_2+\Delta\xi_2)\big) \nonumber\\
		&= \left(4\cos^2\eta\sin^2\frac{\pi}{2m}+4\sin^2\eta\sin^2\frac{\Delta\xi_2}{2}\right)^{\frac{1}{2}}.
	\end{align}
	Note that $n_1=\lfloor2\pi/\Delta\xi_2\rfloor$, with
	\begin{displaymath}
		\Delta\xi_2=2\arcsin\left[\left((d^2/4)\csc^2\eta-\cot^2\eta\sin^2(\pi/2m)\right)^{\frac{1}{2}}\right]
	\end{displaymath}
	 obtained by inverting~\eqref{eq: d displaced circles} and setting the distance to $d$.
	
	Finally, the distance between the image by $\iota_\eta$ of two points aligned with respect to $\xi_1$ in two (alternate) internal circles parametrized by $\xi_2$ and $\xi_2+2\Delta\xi_2$ is obtained by setting $\xi_1' = \xi_1$ and $\xi_2' = \xi_2 + 2\Delta\xi_2$ in \eqref{eq: d-iota-eta-general}:
	\begin{equation} \label{eq: d aligned points}
		d\big(\iota_\eta(\xi_1,\xi_2),\iota_\eta(\xi_1,\xi_2+2\Delta\xi_2)\big) = 2\sin\eta\sin\Delta\xi_2.
		\setcounter{storeeqcounter}{\value{equation}}
	\end{equation}	
	Similarly, we have $n_2=\lfloor2\pi/\Delta\xi_2\rfloor$, with $\Delta\xi_2=\arcsin\big(d/(2\sin\eta)\big)$ obtained from (\ref{eq: d aligned points}) with distance $d$.
	
	As we have to ensure minimum distances both in~\eqref{eq: d displaced circles} and in~\eqref{eq: d aligned points}, we choose the minimum between $n_1$ and $n_2$. Notice moreover that, if we put more than one internal circle, the number of circles~$\tilde{n}$ effectively has to be even, so that first and last circles (which are neighboring circles in the torus $T_{\eta}$) have different displacements, thus ensuring minimum mutual distance between their points (see Fig.~\ref{fig: lattice}).
\end{IEEEproof}

\begin{figure*}[!b]
	\normalsize
	\setcounter{tempeqcounter}{\value{equation}}
	\setcounter{equation}{\value{storeeqcounter}}
	\vspace*{4pt}
	\hrulefill
	\begin{numcases} {M(2n,d) = }
	{\left(M(n,\sqrt{2}d)\right)}^{2} + 2\displaystyle{\sum_{i=1}^{\lfloor t/2 \rfloor}} M(n,d/\cos\eta_i) M(n,d/\sin\eta_i), & for $n>2,$ \label{eq: recursive-1}\\
	m_0n_0 + 2\displaystyle{\sum_{i=1}^{\lfloor t/2 \rfloor}} m_in_i, & for $n=2$ \label{eq: recursive-2}.
	\end{numcases}
	\setcounter{equation}{\value{tempeqcounter}+2}
\end{figure*}

It is possible to describe the generation of points as described above in complex variables, once again referring to the Hopf foliation and noticing that a rotation in $ \mathbb{R}^2 \cong \mathbb{C}$ corresponds to multiplication by a unit complex number. Thus, on each torus $T_\eta$,  points take the form
\begin{displaymath}
	(z_0,z_1) = \big( e^{\boldsymbol{i}(j\Delta\xi_1+k\Delta\xi_1/2)}\cos\eta,\, e^{\boldsymbol{i}(k\Delta\xi_2)}\sin\eta \big),
\end{displaymath}
with $j\in\llbracket 0,m-1\rrbracket$ and  $k\in \llbracket0,n-1\rrbracket$. This description can compare favorably, for instance, to the use of rotation matrices in \cite{torezzan}, because it reduces several matrix products to scalar and complex products. In this work, we have considered the complex description for its simplicity in implementation.

\begin{table*}[!t]
	\renewcommand{\arraystretch}{1.3}
	\centering
	\captionsetup{justification=centering, labelsep=newline}
	\caption{Cardinality of Four-Dimensional Spherical Codes for Different Minimum Distances $d$}
	\label{tab: R4}
	\begin{tabular}{cccccc}
		\hline
		$d$ & SCHF & TLSC \cite{torezzan} & Apple-peeling \cite{elgamal} & Wrapped \cite{hamkins1} & Laminated \cite{hamkins2} \\ \hline \hline
		$0.5$ & $168$ & $172$ & $170$ & * & * \\ 
		$0.4$ & $321$ & $308$ & $342$ & * & * \\ 
		$0.3$ & $774$ & $798$ & $826$ & * & * \\ 
		$0.2$ & $2,683$ & $2,718$ & $2,822$ & * & * \\ 
		$0.1$ & $22,164$ & $22,406$ & $22,740$ & $17,198$ & $16,976$ \\ 
		$0.01$ & $2.27\times 10^7$ & $2.27\times 10^7$ & $1.97\times 10^7$ & ${2.31\times 10^7}^\dagger$ & $2.31\times 10^7$ \\ 
		$0.001$ & $2.27\times 10^{10}$ & $2.27\times 10^{10}$ & $2.27\times 10^{10}$ & ${2.59\times 10^{10}}^\dagger$ & $2.59\times 10^{10}$ \\ \hline
	\end{tabular}\\
	\vspace*{2pt}
	{* unknown values, $\dagger$ estimated values}
\end{table*}

\begin{table*}[!t]
	\renewcommand{\arraystretch}{1.3}
	\centering
	\caption{Cardinality of $n$-Dimensional Spherical Codes for Different Minimum Distances $d$}
	\label{tab: TLSC-higher}
	\begin{tabular}{ccccc}
		\hline
		$n$                   & $d$ & SCHF                  & TLSC ($k$ elements) \cite{naves}              & TLSC (polygon layers) \cite{naves}          \\ \hline \hline
		\multirow{4}{*}{$8$}  & $0.5$ & $4,206$               & $2,748$               & $2,312$  \\  
		& $0.3$ & $150,200$             & $45,252$              & $89,945$             \\  
		& $0.1$ & $3.89 \times 10^{8}$  & $6.47 \times 10^{6}$  & $4.09 \times 10^{8}$ \\
		& $0.01$ & $4.28 \times 10^{15}$  & $7.66 \times 10^{10}$  & $5.19 \times 10^{15}$ \\ \hline
		\multirow{4}{*}{$16$} & $0.5$ & $471,912$             & $69,984$              & $195,312$ \\ 
		& $0.3$ & $2.77 \times 10^{8}$  & $1.17 \times 10^{8}$  & $7.17 \times 10^{7}$ 		\\  
		& $0.1$ & $4.90 \times 10^{15}$ & $2.41 \times 10^{12}$ & $2.39 \times 10^{15}$ 	\\
		& $0.01$ & $6.48 \times 10^{30}$ & $3.66 \times 10^{20}$ & $*$  \\ \hline
		\multirow{4}{*}{$32$} & $0.5$ & $2.47 \times 10^{7}$  & $32$                  & $32,768$   \\  
		& $0.3$ & $4.95 \times 10^{12}$ & $2.68 \times 10^{12}$ & $1.41 \times 10^{12}$ \\  
		& $0.1$ & $1.87 \times 10^{27}$ & $6.81 \times 10^{21}$ & $7.02 \times 10^{24}$ \\ 
		& $0.01$ & $3.96 \times 10^{58}$ & $2.48 \times 10^{38}$ & $*$ \\ \hline
		\multirow{3}{*}{$64$} & $0.5$ & $4.98 \times 10^{9}$  & $64$                  & $2.14 \times 10^{9}$   \\  
		& $0.3$ & $4.61 \times 10^{17}$ & $2.40 \times 10^{11}$ & $9.22 \times 10^{18}$ \\  
		& $0.1$ & $9.35 \times 10^{44}$ & $1.08 \times 10^{38}$ & $2.90 \times 10^{37}$ \\  \hline
	\end{tabular}\\
	\vspace*{2pt}
	{* unknown values}
\end{table*}

\subsection{Recursive Generalization: Spherical Codes in $\mathbb{R}^{2^k}$}

With the generalized foliation of Assertion \ref{assertion:general}, the following natural two-step algorithm for $S^{2n-1}\subset \mathbb{R}^{2n}$ emerges:

\begin{enumerate}
	\item Vary the parameter $\eta\in\left[0,\pi/2\right]$, generating a family of leaves $S_{\cos\eta}^{n-1}\times S_{\sin\eta}^{n-1}$ separated by at least $d$.
	\item On each leaf $S_{\cos\eta}^{n-1}\times S_{\sin\eta}^{n-1}$, distribute points recursively on each of the spheres $S_{\cos\eta}^{n-1}$ and $S_{\sin\eta}^{n-1}$, at scaled minimum distances $d/\cos\eta$ and $d/\sin\eta$, respectively.
\end{enumerate}

We shall focus on dimensions $2^k,\ k\ge2$, starting from $\mathbb{R}^4$. For instance, to construct a spherical code in $S^{15}\subset\mathbb{R}^{16}$, we foliate it by manifolds $\left(S^7\times S^7\right)_{\eta}$, and each copy of $S^7$ is itself foliated by $\left(S^3\times S^3\right)_{\eta'}$. The distribution on each copy of $S^3\subset\mathbb{R}^4$ is known from the basic case.

In our implementation, the standard algorithm exploits in particular the symmetry of the leaves $\left(S^{n-1}\times S^{n-1}\right)_\eta$ around $\eta=\pi/4$. The first chosen leaf  is $\eta_0=\pi/4$, the distribution is done for $\eta \in \left] \pi/4,\pi/2 \right]$ and the points for $\eta \in \left[0,\pi/4\right[$ are obtained by coordinate permutations.

As a direct result of the proposed construction, we immediately derive the following Proposition~\ref{prop:cardinality}:

\begin{proposition} 
	\label{prop:cardinality}
	The cardinality $M(2n,d)$ of a SCHF in dimension~$2n$, with minimum distance~$d$, constructed by our standard procedure, is given by the recursive expressions \eqref{eq: recursive-1} and \eqref{eq: recursive-2} at the bottom of the page,	with $\eta_i \coloneqq \pi/4+i\Delta\eta$ as in \eqref{eq: Deta}, $t \coloneqq t(d)$ as in \eqref{eq: t}, $m_i \coloneqq m(d,\eta_i)$ as in \eqref{eq: MNP on a circle m} and $n_i \coloneqq n(d,\eta_i)$ as in \eqref{eq: MN circles n}.
\end{proposition}

\begin{example}
	To construct a code in $\mathbb{R}^8$ with minimum distance $d=0.5$ by our standard procedure, we consider the foliation of $S^7$ by $\left(S^3 \times S^3\right)_\eta$ and use Proposition \ref{prop:dist-tori} to choose the set of parameters $\eta\in\{0.2800, \pi/4, 1.2908\}$. Next, for each leaf $S^3_{\cos\eta} \times S^3_{\sin\eta}$, we take the Cartesian product of codes in the $3$-spheres of radii $\cos\eta$ and $\sin\eta$ in $\mathbb{R}^4$. On each of these $3$-spheres, we apply the basic-case algorithm for minimum distances $d/\cos\eta$ and $d/\sin\eta$, namely: choose a family of tori $T_\eta$ and distribute points on each. For instance, $\left(S^3\times S^3\right)_{1.2908} = S^3_{0.2764} \times S^3_{0.9610}$. On the first-component sphere, it is only possible to choose one torus, with $\eta = \pi/4$. On the second-component sphere, we choose the tori with $\eta\in\{0.2591, \pi/4, 1.3117\}$.  Due to the symmetry about $\eta = \pi/4$, in both cases, it suffices to calculate half of the points and obtain the symmetric ones by permuting their coordinates. Summing across all the leaves, there are 2,608 points in total.
\end{example}

\subsection{Modifications} \label{sec: modifications}

One can consider some small modifications of the previous standard procedure, in order to improve the cardinality of the code.

\begin{enumerate}
	\item When choosing leaves, in the context of Corollary \ref{cor: t}, we may choose not only symmetrically distributed leaves around $\eta=\pi/4$, but consider the following choices -- and even a combination of them -- in different dimensions:
	\begin{enumerate}
		\item $\eta = \pi/4\pm k\Delta\eta$, for $k\in\llbracket0,\lfloor t(d)/2\rfloor\rrbracket$;
		\item $\eta = k\Delta\eta$, for $k\in\llbracket0, t(d)\rrbracket$;
		\item $\eta = \pi/2-k\Delta\eta$, for $k\in\llbracket0, t(d)\rrbracket$.
	\end{enumerate}
	
	\item When distributing points on a torus $T_\eta$, in the context of Proposition \ref{prop:T_eta}, we may consider only the `diagonal' internal circles, i.e., the images $\iota_\eta(\xi_1, \xi_2)$ with $\xi_1=\xi_2$, whenever that is more advantageous than the standard distribution. As those circles have unit radius, the number of points that can be placed on them, respecting minimum mutual distance $d$, is $\lfloor\pi/\arcsin(d/2)\rfloor$.
	
	\item Whenever possible and more advantageous, we can consider explicit \emph{ad hoc} constructions \cite{ericson}: optimal codes in $\mathbb{R}^4$ for cardinalities 
	$2$, $3$, $4$, $5$, $8$, $10$ and $24$ (minimum distances $2$, $\sqrt{3}$, $\sqrt{8/3}$, $\sqrt{5/2}$, $\sqrt{2}$, $\sqrt{5/3}$, $1$, respectively), as well as the biorthogonal codes which place $2n$ points with minimum distance $\sqrt{2}$ in any dimension~$n$.
\end{enumerate}

\begin{remark}
	In the proposed standard procedure, any dimension $n$ can be considered as a basic case for codes in $\mathbb{R}^{2n}$, so long as good constructive codes are available in $\mathbb{R}^{n}$ for a wide range of minimum distances. However much this may provide  greater density, as discussed in Section~\ref{sec:density}, in this paper we focus on the construction in $\mathbb{R}^{2^k}$ with basic case $\mathbb{R}^{4}$ due to its effective constructiveness and low complexity.
\end{remark}

\begin{table}[!t]
	\renewcommand{\arraystretch}{1.3}
	\centering
	\caption{Cardinality of $n$-Dimensional Spherical Codes for Different Minimum Distances $d$}
	\label{tab: EQPA}
	\begin{tabular}{cccc}
		\hline
		$n$                   & $d$      & SCHF                 & EQ codes \cite{leopardi}     \\  \hline \hline
		\multirow{3}{*}{$4$}  & $0.27944$ & $918$                & $500$    \\ 
		& $0.23707$ & $1,540$              & $1,000$  \\ 
		& $0.10374$ & $19,768$             & $10,000$ \\ \hline
		\multirow{3}{*}{$8$}  & $0.51282$ & $3,228$              & $500$    \\
		& $0.47025$ & $5,889$              & $1,000$  \\
		& $0.31379$ & $100,074$            & $10,000$ \\ \hline
		\multirow{3}{*}{$16$} & $0.56498$ & $23,882$             & $500$    \\  
		& $0.51483$ & $182,424$            & $1,000$  \\
		& $0.40868$ & $2.06 \times 10^{6}$ & $10,000$ \\ \hline
		\multirow{3}{*}{$32$} & $0.45847$ & $4.07 \times 10^{7}$ & $500$    \\  
		& $0.44805$ & $1.68 \times 10^{8}$ & $1,000$  \\ 
		& $0.41207$ & $6.59 \times 10^{8}$ & $10,000$ \\ \hline
	\end{tabular}
\end{table}

\begin{table}[!t]
	\renewcommand{\arraystretch}{1.3}
	\caption{Cardinality of Four-Dimensional Spherical Codes for Different Minimum Distances $d$}
	\label{tab: group-codes}
	\centering
	\begin{tabular}{ccc}
		\hline
		$d$        & SHCF    & CGC \cite{siqueira} \\ \hline \hline
		$0.330158$ & $556$   & $200$                  \\ 
		$0.237033$ & $1,586$ & $400$                  \\ 
		$0.193059$ & $2,988$ & $600$                  \\ 
		$0.16806$  & $4,535$ & $800$                  \\ 
		$0.149405$ & $6,450$ & $1,000$                \\ \hline
	\end{tabular}
\end{table}

\begin{table}[!t]
	\renewcommand{\arraystretch}{1.3}
	\caption{Cardinality of $n$-Dimensional Spherical Codes for Different Minimum Distances $d$}
	\label{tab: group-codes-2}
	\centering
	\begin{tabular}{cccc}
		\hline
		$n$ & $d$        & SCHF      & CGC \cite{alves} \\ \hline \hline
		\multirow{4}{*}{4}
		& $0.012706$   & $11,067,004$  & $141,180$  \\  
		& $0.00733585$ & $57,610,534$  & $423,540$  \\ 
		& $0.00465076$ & $226,265,570$ & $1,053,780$ \\ 
		& $0.00423537$ & $299,595,092$ & $1,270,620$ \\ \hline
		\multirow{4}{*}{8}
		& $0.707107$   &   $416$      &   $648$    \\  
		& $0.541196$   & $2,342$      & $2,048$    \\ 
		& $0.437016$   & $9,700$      & $5,000$    \\
		& $0.366025$   & $38,244$     & $10,368$   \\ \hline
	\end{tabular}
\end{table}

\section{Non-asymptotic Performance Analysis} \label{sec:performance}

We compare the cardinality of our codes with other constructive spherical codes, in different dimensions, for many non-asymptotic minimum distance regimes. In dimension~4, we compare our results with apple-peeling\footnote{We follow the description in \cite{hamkins1}, using the implementation generously shared by its authors.}~\cite{elgamal}, wrapped~\cite{hamkins1}, laminated~\cite{hamkins2} and the torus layers spherical codes~(TLSC)~\cite{torezzan} (Table~\ref{tab: R4}). In higher dimensions, we compare with two TLSC implementations by Naves~\cite{naves} that differ in the choice of the subcode: either with $k$ elements or on polygon layers (Table~\ref{tab: TLSC-higher}). In these dimensions, we have also considered codes generated by the equal area sphere partitioning algorithm~(EQ codes)~\cite{leopardi} (Table \ref{tab: EQPA}), concatenated MPSK~\cite[p. 36]{hamkins3}, commutative group codes~(CGC)~\cite{siqueira, alves} (Tables~\ref{tab: group-codes} and~\ref{tab: group-codes-2}) and some codes from Sloane \emph{et al.}~\cite{sloane-table}. The SCHF considered in these tables use the modifications introduced in Section~\ref{sec: modifications}. The proposed SCHF construction was implemented in \emph{Wolfram Mathematica} and \emph{Python}.

We see that the performance of SCHF in $\mathbb{R}^4$ is, as expected, similar to TLSC and not far from some of the best known spherical codes. In higher dimensions, SCHF can achieve a higher cardinality than TLSC~($k$ elements) in most regimes and, in many of them, higher than TLSC~(polygon layers) too. Commutative group codes, which have a powerful algebraic structure, are outperformed by SCHF in nearly all considered minimum distance regimes.

\begin{table}[!t]
	\renewcommand{\arraystretch}{1.3}
	\centering
	\caption{Cardinality of Four-Dimensional Spherical Codes for Different Minimum Distances $d$}
	\label{tab: compare-rodrigues}
	\begin{tabular}{ccc}
		\hline
		$d$        & SCHF  & Rodrigues \emph{et al.} \cite{rodrigues} \\ \hline \hline
		$0.488876$ & $174$ & $112$        \\
		$0.389872$ & $344$ & $128$        \\ \hline
	\end{tabular}
\end{table}

\begin{figure*}[!t]
	\centering
	\includegraphics[height=5cm]{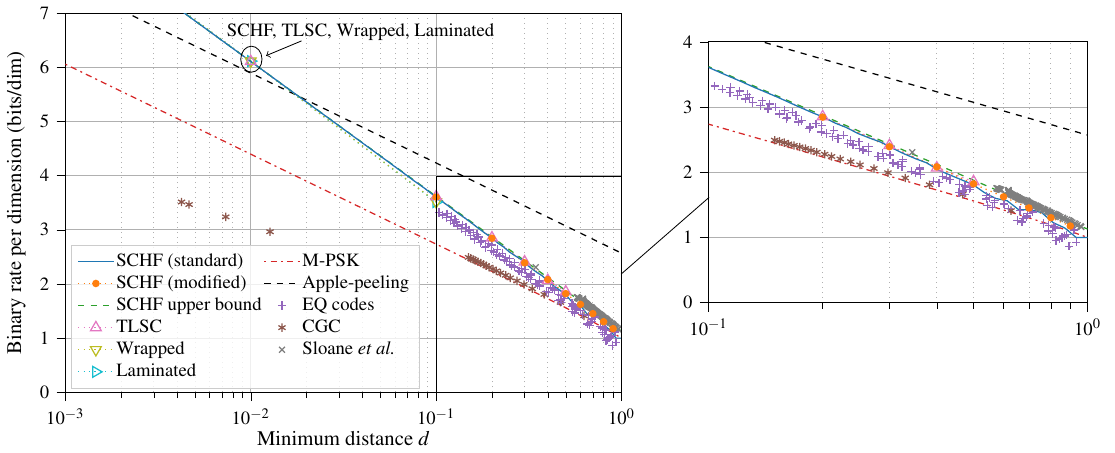}
	\caption{Binary rate per dimension for different codes in dimension 4 with detail.}
	\label{fig:plot4}
\end{figure*}

\begin{figure}[!t]
	\centering
	\includegraphics[height=5cm]{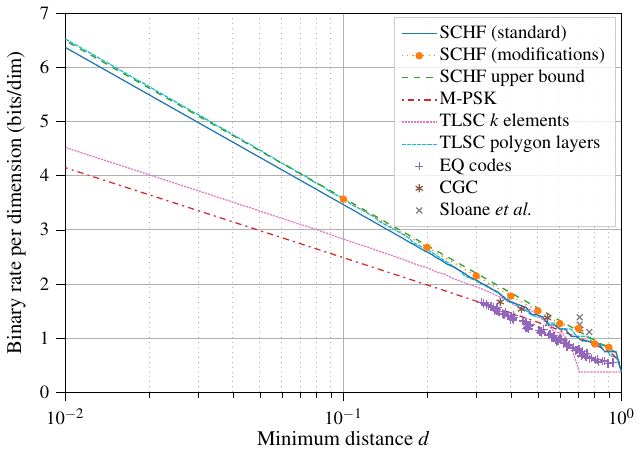}
	\caption{Binary rate per dimension for different codes in dimension 8.}
	\label{fig:plot8}
\end{figure}

\begin{figure}[!t]
	\centering
	\includegraphics[height=5cm]{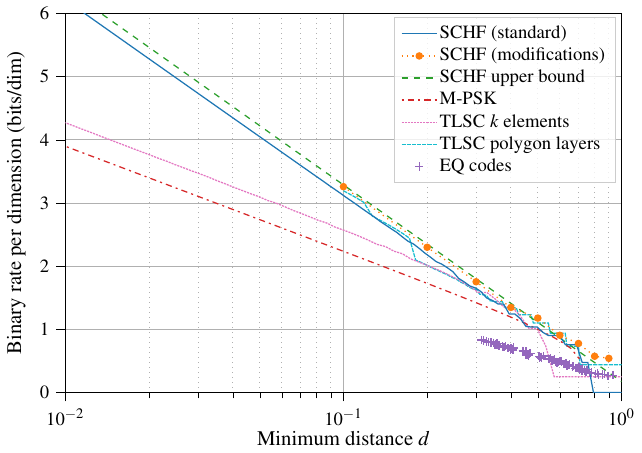}
	\caption{Binary rate per dimension for different codes in dimension 16.}
	\label{fig:plot16}
\end{figure}

\begin{figure}[!t]
	\centering
	\includegraphics[height=5cm]{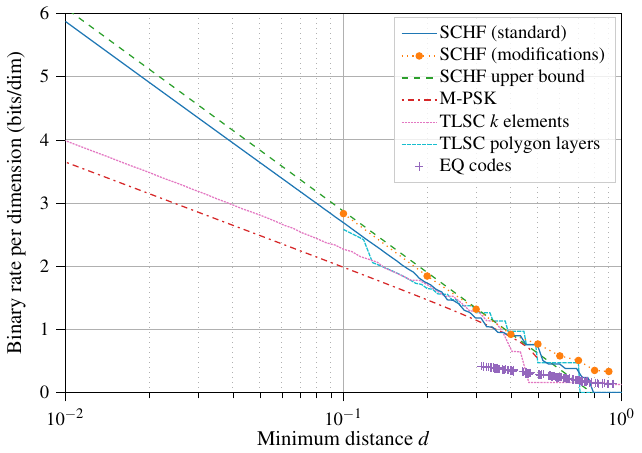}
	\caption{Binary rate per dimension for different codes in dimension 32.}
	\label{fig:plot32}
\end{figure}

A more complete picture is given on Fig.~\ref{fig:plot4}, \ref{fig:plot8}, \ref{fig:plot16}, \ref{fig:plot32}, showing the binary rate per dimension $R = (\log_2 M)/n$ for codes in dimensions 4, 8, 16 and 32, respectively. These computations for the proposed recursive SCHF, both with and without the modifications of Section~\ref{sec: modifications}, show in each dimension and for small values of~$d$ a good approximation of the asymptotic bounds derived in Section~\ref{sec:density}. Indeed, SCHF generally outperforms the other plotted constructions.

We also acknowledge the recent appearance, on a somewhat different vein, of some Hopf fibration formalism in the context of optical communications~\cite{rodrigues}. A design for higher order modulations is presented, based on an interesting use of the Hopf preimage under the so-called \emph{sampled discrete Hopf fibration} and in close relation to physical properties of light. That construction relies on the choice of a polytope on the base space (such as the tetrakis hexahedron) and apparently does not address the spherical packing problem for any given minimum distance. Particularities aside, SCHF outperform the two four-dimensional modulations presented in~\cite{rodrigues} at the same minimum distance, cf. Table~\ref{tab: compare-rodrigues}.

\section{Asymptotic Density} \label{sec:density}

We now analyze the density of our spherical codes. Consider the gamma function $\Gamma(x)=\int_{0}^{\infty} e^{-t}t^{x-1}~\mathrm{d}t$. The Euclidean $\left(n-1\right)$-dimensional volume (hypersurface area) of the sphere $S^{n-1}\subset\mathbb{R}^n$ is given~\cite{conway} by
\begin{displaymath}
	\mathbb{S}_n \coloneqq \frac{n\pi^{n/2}}{\Gamma(1 + n/2)},
\end{displaymath}
and the corresponding $n$-dimensional volume of the ball bounded by $S^{n-1}$ is
\begin{displaymath}
	\mathbb{V}_n \coloneqq \frac{\pi^{n/2}}{\Gamma(1 + n/2)}.
\end{displaymath}
Note that the spherical code with minimum distance $d$ has minimum angular separation $\theta(d) = 2\arcsin(d/2)$. The $\left(n-1\right)$-dimensional volume of a spherical cap on the sphere $S^{n-1}$ with angular radius $\theta(d)/2$ is
\begin{displaymath}
	\mathbb{S}_C(n,d) \coloneqq \mathbb{S}_{n-1}\int_{0}^{\theta(d)/2} \sin^{n-2}x~\mathrm{d}x.
\end{displaymath}
Hence the density $\Delta(\mathcal{C})$ of a $n$-dimensional spherical code $\mathcal{C}(M,n,d)$ is the ratio of the total area covered by the $M(n,d)$ spherical caps, with angular radius $\theta(d)/2$ centered at the codewords, by the total surface area:
\begin{equation} \label{eq: sc-density}
	\Delta(\mathcal{C}(n,d)) \coloneqq \frac{M(n,d) \, \mathbb{S}_C(n,d)}{\mathbb{S}_n}.
\end{equation}

In what follows, we will write $f(d) \simeq g(d)$ when
\begin{displaymath}
	\lim_{d\to 0} \frac{f(d)}{g(d)} = 1.
\end{displaymath}
For small values of $d$, $\mathbb{S}_C(n,d)$ can be approximated \cite{hamkins1} by
\begin{displaymath}
	\mathbb{S}_C(n,d) = \mathbb{V}_{n-1}\left(\frac{d}{2}\right)^{n-1}+O(d^{n+1}),
\end{displaymath}
which implies $\mathbb{S}_C(n,d) \simeq \mathbb{V}_{n-1} \left(\frac{d}{2}\right)^{n-1}$ and
\begin{equation}  \label{eq: approx-small-d}
	\Delta(\mathcal{C}(n,d)) \simeq \frac{M(n,d) \mathbb{V}_{n-1}}{\mathbb{S}_n} \left(\frac{d}{2}\right)^{n-1}.
\end{equation}

The \emph{center density} of a spherical code $\mathcal{C}(M,n,d)$ is defined as $\Delta_c(\mathcal{C}(n,d)) \coloneqq \Delta(\mathcal{C}(n,d))/\mathbb{V}_{n-1}$ and its asymptotic value for a family of spherical codes, constructed for different minimum distances~$d$, is
\begin{equation} 
\label{eq: def-centre-density}
\overline{\Delta}_c (\mathcal{C}(n)) \coloneqq \lim_{d \to 0} \frac{M(n,d)}{\mathbb{S}_{n}} \left(\frac{d}{2}\right)^{n-1}.
\end{equation}
It provides a means of comparing packings of different constructions in a given dimension, for small $d$.

\begin{lemma} 
	\label{lemma: SCHF-density-R4}
	The asymptotic center density of the SCHF in dimension $4$ is
	\begin{equation}
	\overline{\Delta}_c ({\mathrm{SCHF}}[4]) = \frac{1}{4\sqrt{3}}.
	\end{equation}
\end{lemma}
\begin{IEEEproof}
	In $\mathbb{R}^4$, the asymptotic SCHF and TLSC coincide, and their densities can be approached by the density of the lattice product $A_2 \times \mathbb{Z}$~\cite[Proposition 6]{torezzan}. Considering the well-known center densities of lattices $A_2$ and $\mathbb{Z}$~\cite{conway}, we then have:
	\begin{align*}
	\overline{\Delta}_c (\mathrm{SCHF}[4])
	= \Delta_c (A_2 \times \mathbb{Z})
	&= \Delta_c (A_2) \cdot \Delta_c (\mathbb{Z})\\
	&= \frac{1}{2\sqrt{3}} \cdot \frac{1}{2}
	= \frac{1}{4\sqrt{3}}.
	\end{align*}
\end{IEEEproof}

\begin{proposition} \label{prop: asymptotic-density}
	The asymptotic center density of the SCHF in dimension $2n$, constructed from a family of codes $\mathcal{C}(n)$ in dimension $n$, which achieves asymptotic density $\overline{\Delta}_c(\mathcal{C}(n))$, is
	\begin{displaymath}
	\overline{\Delta}_c(\mathrm{SCHF}[2n;\mathcal{C}(n)]) = \frac{1}{2} {\left(\overline{\Delta}_c (\mathrm{\mathcal{C}}(n)) \right)}^{2}.
	\end{displaymath}
\end{proposition}

\begin{IEEEproof}
	For small $d$, we have
	\begin{equation} \label{eq: Dc-2n}
		\overline{\Delta}_c(\mathrm{SCHF}[2n;\mathcal{C}(n)]) \simeq \frac{M(2n,d)}{\mathbb{S}_{2n}} {\left(\frac{d}{2}\right)}^{2n-1}
	\end{equation}
	and
	\begin{equation} \label{eq: Dc-n}
		\overline{\Delta}_c(\mathrm{\mathcal{C}}(n)) \simeq \frac{M(n,d)}{\mathbb{S}_{n}} {\left(\frac{d}{2}\right)}^{n-1}.
	\end{equation}
	From \eqref{eq: Dc-n}, we have
	\begin{equation} \label{eq: M-nd}
		M(n,d) \simeq \mathbb{S}_n {\left(\frac{2}{d}\right)}^{n-1} \overline{\Delta}_c(\mathcal{C}(n)).
	\end{equation}
	In the asymptotic behavior, the particular choice of leaves as introduced in Section~\ref{sec: modifications} is irrelevant. Therefore, we consider, for simplicity, the leaves $\eta_i = i\Delta\eta$.
	From the construction of SCHF, we have	
	\begin{IEEEeqnarray*}{rCll}
		M(2n,d) &=& \IEEEeqnarraymulticol{2}{l}{
			\sum_{i=0}^{t(d)} M(n, d/\cos\eta_i) M(n, d/\sin\eta_i)
		}\\
		&\simeq& \sum_{i=0}^{t(d)} & \Bigg[ \left( \mathbb{S}_n \left(\frac{2\cos\eta_i}{d}\right)^{n-1} \overline{\Delta}_c(\mathcal{C}(n)) \right)
		\\
		&\quad& &\times\left( \mathbb{S}_n \left(\frac{2\sin\eta_i}{d}\right)^{n-1} \overline{\Delta}_c(\mathcal{C}(n)) \right) \Bigg] \\
		&=& \IEEEeqnarraymulticol{2}{l}{
			[\overline{\Delta}_c(\mathcal{C}(n)]^{2} (\mathbb{S}_n)^{2} \frac{2^{n-1}}{d^{2n-2}} \sum_{i=0}^{t(d)} {\left[\sin (2\eta_i)\right]}^{n-1}
		}.
	\end{IEEEeqnarray*}
	Therefore, from \eqref{eq: Dc-2n},
	\begin{align*}
		\hspace{4em}&\hspace{-4em}\overline{\Delta}_c(\mathrm{SCHF}\left[2n;\mathcal{C}(n)\right])\\
		&\simeq \frac{[\overline{\Delta}_c(\mathcal{C}(n)]^2}{2^n} \frac{(\mathbb{S}_n)^{2}}{\mathbb{S}_{2n}} \sum_{i=0}^{t(d)}  \left[\sin (2\eta_i)\right]^{n-1} d.
	\end{align*}
	
	For small $d$, we have $\Delta\eta \simeq d$, $\eta_i \simeq id$ and $t(d) \simeq \pi/2d$, hence the last summation approaches the corresponding integral, which implies
	\begin{align} \label{eq: integral}
		\hspace{4em}&\hspace{-4em}\overline{\Delta}_c\left(\mathrm{SCHF}[2n;\mathcal{C}(n)]\right) \nonumber\\
		&= \frac{[\overline{\Delta}_c(\mathcal{C}(n)]^2}{2^n} \frac{(\mathbb{S}_n)^2}{\mathbb{S}_{2n}} \int_{0}^{\pi/2} {(\sin 2\eta )}^{n-1}~\mathrm{d}\eta \nonumber\\
		&= \frac{[\overline{\Delta}_c(\mathcal{C}(n)]^2}{2^{n+1}} \frac{(\mathbb{S}_n)^2}{\mathbb{S}_{2n}} \int_{0}^{\pi} {(\sin x )}^{n-1}~\mathrm{d}x.
	\end{align}
	
	The sphere $S^{2n-1}$ is foliated by the leaves $S^{n-1}_{\cos\eta_i} \times S^{n-1}_{\sin\eta_i}$, with $\eta_i \in \left[0, \pi/2\right]$, and the distance between the leaves is the chordal distance in $S^1$~(Proposition~\ref{prop:dist-tori}). Hence, for small $d$,  the arc-chord approximation $d \simeq \Delta \eta$ yields
	\begin{align}
		\mathbb{S}_{2n} &\simeq \sum_{i=0}^{t(d)}\mathsf{Vol}(S^{n-1}_{\cos\eta_i} \times S^{n-1}_{\sin\eta_i}) \Delta\eta \nonumber\\
		&= \sum_{i=0}^{t(d)} \mathsf{Vol}(S^{n-1}_{\cos\eta_i}) \mathsf{Vol}(S^{n-1}_{\sin\eta_i}) \Delta\eta \nonumber \\	
		&= \sum_{i=0}^{t(d)} \mathbb{S}_n{(\cos\eta_i)}^{n-1} \, \mathbb{S}_n{(\sin\eta_i)}^{n-1} \, \Delta\eta \nonumber \\
		&= \frac{(\mathbb{S}_n)^2}{2^{n-1}} \sum_{i=0}^{t(d)} \left[\sin(2\eta_i)\right]^{n-1} \Delta\eta \nonumber,
	\end{align}
	where $\mathsf{Vol}(\cdot)$ denotes the volume of the object, and, when $d \to 0$, as in \eqref{eq: integral},
	\begin{equation} \label{eq: S-2n}
	\mathbb{S}_{2n} = \frac{(\mathbb{S}_n)^2}{2^{n}} \int_{0}^{\pi} (\sin x)^{n-1}~\mathrm{d}x.
	\end{equation}
	
	Substituting \eqref{eq: S-2n} in \eqref{eq: integral} yields the claim:
	\begin{equation}
		\overline{\Delta}_c\left(\mathrm{SCHF}[2n;\mathcal{C}(n)]\right) = \frac{1}{2} {\left(\overline{\Delta}_c (\mathrm{\mathcal{C}}(n)) \right)}^{2}.
	\end{equation}
\end{IEEEproof}

\begin{remark} \label{remark: SCHF-density-half}
	Since the maximum asymptotic center density for spherical codes in $S^{n-1} \subset \mathbb{R}^n$ is the highest center packing density of $\mathbb{R}^{n-1}$, denoted by ${\Delta}_c(\mathbb{R}^{n-1})$ (cf. Proposition~\ref{prop: asymptotic-density}), the asymptotic density of a SCHF in $S^{2n-1} \subset \mathbb{R}^{2n}$ is  bounded above and asymptotic to
	\begin{equation}
		{\overline{\Delta}}_c\left(\mathrm{SCHF}[2n; \mathcal{C}(n)]\right) \le \frac{1}{2} {\left(\Delta_c (\mathbb{R}^{n-1})\right)}^{2}.
	\end{equation}
\end{remark}

By recursively applying Proposition~\ref{prop: asymptotic-density} in dimensions $2^{k}$, for $k \ge 2$ (cf. Lemma~\ref{lemma: SCHF-density-R4}), we get

\begin{corollary} 
	\label{cor: SCHF-density-2k}
	The asymptotic center density  of the recursive SCHF in dimension $n=2^k$, $k \ge 2$ is
	\begin{align} \label{eq: density}
	\overline{\Delta}_c (\mathrm{SCHF}[2^k]) &\coloneqq \overline{\Delta}_c \big(\mathrm{SCHF}[2^k; \mathrm{SCHF}[2^{k-1}; \cdots \mathrm{SCHF}[4]]]\big) \nonumber\\
	&\;= {(2)}^{1-(3){2}^{k-2}} {(3)}^{-{2}^{k-3}}.
	\end{align}
\end{corollary}

Moreover, using \eqref{eq: M-nd} and Corollary~\ref{cor: SCHF-density-2k}, we get:

\begin{corollary} \label{cor: SCHF-M-2k}
	The cardinality $M(n,d)$ of the recursive SCHF in dimension $n=2^k$ is bounded above and, as $d \to 0$, asymptotic to
	\begin{eqnarray} 
	\label{eq: SCHF-M-2k}
	\overline{M}(2^k,d) = \frac{ {2}^{k+{2}^{k-2}}~{3}^{-{2}^{k-3}}~{\pi}^{{2}^{k-1}} }{ ({2}^{k-1})!~{d}^{{2}^{k}-1} }.
	\end{eqnarray}
\end{corollary}

\begin{IEEEproof}
	For small values of $d$, we have the approximation
	\begin{displaymath}
		\overline{\Delta}_c \simeq \frac{M(n,d)}{\mathbb{S}_n}\left(\frac{d}{2}\right)^{n-1}.
	\end{displaymath}
	Using \eqref{eq: density}, we get
	\begin{align*}
		\overline{M}(2^k,d) &= \overline{\Delta}_c(\mathrm{SCHF}[2^k]) ~\mathbb{S}_{2^k} \left(\frac{2}{d}\right)^{2^k-1}\\
		&= \left({(2)}^{1-(3){2}^{k-2}} {(3)}^{-{2}^{k-3}}\right) \frac{2^k \pi^{2^{k-1}}}{(2^{k-1})!} \left(\frac{2}{d}\right)^{2^k-1}\\
		&= \frac{ {2}^{k+{2}^{k-2}}~{3}^{-{2}^{k-3}}~{\pi}^{{2}^{k-1}} }{ ({2}^{k-1})!~{d}^{{2}^{k}-1} }. \qedhere
	\end{align*}
\end{IEEEproof}

\begin{table*}[!t]
	\renewcommand{\arraystretch}{2.8}
	\centering
	\caption{Asymptotic Center Density for Different $n$-Dimensional Spherical Codes. $\Delta_{n}$ denotes the highest center density  of a sphere in $\mathbb{R}^n$: $\Delta_n \coloneqq {\Delta}_c (\mathbb{R}^{n-1})$. For dimensions 8, 16 and 32 the calculations assume the best known center densities.}
	\label{tab: density}
	\begin{tabular}{cAAAAAAAAAA}
		\hline
		& \multicolumn{2}{c}{SCHF (recursive)} & \multicolumn{2}{c}{SCHF (half-dimension)} & \multicolumn{2}{c}{TLSC} & \multicolumn{2}{c}{Apple-peeling} & \multicolumn{2}{c}{$n-1$ packing} \\ \hline \hline
		
		$n$ & \multicolumn{2}{c}{${\left(2^{\frac{3n}{4}-1}3^{\frac{n}{8}}\right)}^{-1}$} & \multicolumn{2}{c}{$\dfrac{1}{2}\left(\Delta_{\frac{n}{2}-1}\right)^2$} & \multicolumn{2}{c}{$\Delta_{\frac{n}{2}}\Delta_{\frac{n}{2}-1}$} & \multicolumn{2}{c}{\scriptsize$\dfrac{\mathbb{V}_{n-2}}{\mathbb{V}_{n-1}}\dfrac{\Delta_{n-2}}{2}\beta\left(\frac{n}{2},\frac{1}{2}\right)$} & \multicolumn{2}{c}{$\Delta_{n-1}$} \vspace{0.5mm} \\ \hline
		
		$4$ & \dfrac{1}{4\sqrt{3}} &\approx 0.1443 & &- & \dfrac{1}{4\sqrt{3}} &\approx 0.1443 & \dfrac{1}{4\sqrt{3}} &\approx 0.1443 & \dfrac{1}{4 \sqrt{2}} &\approx 0.1768 \\
		
		$8$  & \dfrac{1}{96} &\approx 0.0104 & \dfrac{1}{64} &\approx 0.0156 & \dfrac{1}{32\sqrt{2}} &\approx 0.0221 & \dfrac{1}{16\sqrt{3}} &\approx 0.03608 & \dfrac{1}{16} &= 0.0625 \\ 		

		$16$ & \dfrac{1}{18,432} &\approx 5.42 \times 10^{-5} & \dfrac{1}{512} &\approx 0.0020 & \dfrac{1}{256} &\approx 0.0039 & \dfrac{1}{32\sqrt{3}} &\approx 0.0180 & \dfrac{1}{16 \sqrt{2}} &\approx 0.0442 \\
		
		$32$ & \dfrac{1}{679,477,248} &\approx 1.47 \times 10^{-9} & \dfrac{1}{1,024} &\approx 0.0010 & \dfrac{1}{256\sqrt{2}} &\approx 0.0028 & \dfrac{3^{13}\sqrt{3}}{2^{23}} &\approx 0.3292 & \dfrac{3^{15}}{2^{23.5}} &\approx 1.2095 \vspace{0.5mm} \\ \hline
	\end{tabular}\\
	\vspace*{-3pt}
\end{table*}

For a fixed dimension~$n$, the asymptotic center density  of different spherical code constructions allows one to compare their respective numbers of codewords for the same small minimum distance~$d$, cf.~\eqref{eq: M-nd}.

Table~\ref{tab: density} compares some asymptotic center densities for spherical codes in dimensions 4, 8, 16 and 32. We consider both the SCHF recursive procedure with basic case $\mathbb{R}^4$ (Corollary~\ref{cor: SCHF-density-2k}) and the SCHF procedure that uses asymptotic dense codes in the half dimension (Remark~\ref{remark: SCHF-density-half}). For each dimension, we also include the TLSC upper bound as in~\cite[Proposition 6]{torezzan}, the apple-peeling bound as in~\cite[Lemma 3]{hamkins1} and highest asymptotic center density  for spherical codes (from the best known packing in the previous dimension)~\cite{conway, nebe}.

We can see that the ratios between the center densities of these different constructions show how much smaller the number of codewords achieved by recursive SCHF construction is, when compared SCHF using half-dimension codes, TLSC, apple-peeling and, of course, the highest possible asymptotic density in each dimension, which can be theoretically achieved by wrapped or laminated codes. The trade-off to emphasize here is the high constructibility of recursive SCHF for any given minimum distance and its low complexity in the encoding and decoding processes, which will be discussed in Sections~\ref{sec:complexity} and~\ref{sec:decoding}.

One should also point out that there may be a difference between asymptotic density bounds and the density effectively achieved, especially for higher dimensions. This is due to the characteristics of each of the analyzed constructions. Half-dimension SCHF depend on the existence of good codes in the half dimension; TLSC require the use of the best codes and lattices in the half dimension; wrapped codes rely on the choice of a lattice in the previous dimension; laminated codes have been approached in dimensions from 2 to 49 and may have slower convergence than wrapped codes; apple-peeling construction is based on spherical codes in the previous dimension. We note that, in the case of TLSC, the implementations carried out so far do not seek to construct the densest theoretically possible codes, but rather good, feasible ones -- as in~\cite{naves}, which proposes different approaches for the construction of the subcodes. On the other hand, in dimensions~$2^k$, the construction of recursive SCHF does not depend on any choice, can be done for any given minimum distance~$d$ and the asymptotic bound is indeed approached in the results shown in Fig.~\ref{fig:plot4}, \ref{fig:plot8}, \ref{fig:plot16}, \ref{fig:plot32}.

The construction of SCHF using better available constructions in the half dimension offers an easy way of obtaining good codes. One could consider, for instance, using a family of wrapped codes in dimension 25 (as  in~\cite{hamkins4}, based on the Leech lattice) to construct codes in dimension 50 by  Hopf foliations, with low addition in encoding complexity, cf.~Section \ref{sec:complexity} (a wrapped spherical code in this dimension would require the use of a good lattice in dimension 49).

Finally, it is also interesting to compare the asymptotic behavior of recursive SCHF with the more structured spherical commutative group codes~(CGC). From~\cite[Proposition 7]{siqueira}, the number  $M(n,d)$ of codewords of a CGC in dimension $n$ is bounded above by
\begin{equation}
	M(n,d) < \Delta_c(\Lambda_{n/2}) {\left( \frac{4\pi}{d \sqrt{n/2}} \right)}^{n/2},
\end{equation}
where $\Delta_c(\Lambda_{n/2})$ is the maximum center density of a lattice packing in $\mathbb{R}^{n/2}$. This implies that the asymptotic center density is equal to zero, which is expected, since those codes must be contained in a $n$-dimensional flat torus. Note that, for a fixed dimension $n=2^k$, the cardinality of a CGC grows with $O(1/d^{2^{k-1}})$, while for a recursive SCHF (cf. Corollary~\ref{cor: SCHF-M-2k}), it grows with $O(1/d^{2^{k}-1})$, i.e., there exists a value~$d$ beyond which the cardinality of SCHF outperforms CGC (see Tables~\ref{tab: group-codes} and~\ref{tab: group-codes-2}).

\section{Encoding Complexity Analysis} \label{sec:complexity}

We now present a complexity analysis  of the encoding algorithm for the standard SCHF construction. Lachaud and Stern~\cite{lachaud} propose the following definition for polynomial complexity of a spherical code. Let $\Sigma$ be a finite alphabet and consider spherical codes $\mathcal{C}=\mathcal{C}(M,n,d)\subset S^{n-1}$ that are images of maps $F:\Sigma^k \to S^{n-1}$. We say that a family $(\mathcal{C}_i)$ of spherical codes is \textit{polynomially constructible} if there is a sequence $(F_i)$ of maps $F_i:\Sigma^{k_i} \to S^{n_i-1}$ such that: (i)~$F_i$ is one-to-one from $\Sigma^{k_i}$ to $\mathcal{C}_i$, and (ii)~for every $a\in\Sigma^{k_i}$, the point $F_i(a)$ is computable from $i$ and $a$ in polynomial time, with respect to the dimension $n_i$ of $\mathcal{C}_i$.

\subsection{Basic Case: Spherical Codes in $\mathbb{R}^4$}

We first analyze the encoding complexity in dimension $n=4$. The injection $F$ can be decomposed as $F=\iota \circ \chi$, where $\iota$ is as in~\eqref{eq: iota S^3 por T^2} and $\chi(a)=(\eta;\xi_1,\xi_2)$ as in Algorithm~\ref{alg-enc-4}. We assume that, in the construction of the code $\mathcal{C}(M,4,d)$, we store a table that contains information on each leaf $T_\eta$: each line contains the index $i$ of the leaf, the parameter $\eta_i$ and the number of points in that leaf $M_i$. The length of this table is $t(d) = \left\lfloor \pi/4\arcsin(d/2) \right\rfloor$ (Corollary~\ref{cor: t}), hence the storage complexity is $O(t)$. We assume that accessing data in this table has constant complexity.

\begin{algorithm}[!t]
	\caption{Encoding algorithm for $\mathbb{R}^4$ (map $F=\iota\circ\chi$).}
	\label{alg-enc-4}
	\begin{algorithmic}[1]
		\renewcommand{\algorithmicrequire}{\textbf{Input:}}
		\renewcommand{\algorithmicensure}{\textbf{Output:}}
		\REQUIRE $a$, $d$
		\ENSURE $\mathbf{x} = (x_1,x_2,x_3,x_4)$
		
		\STATE $t \leftarrow \left\lfloor \pi/ 4\arcsin(d/2) \right\rfloor$
		\FOR {$i\in\llbracket -\lfloor t/2 \rfloor,\lfloor t/2 \rfloor \rrbracket$}
		\STATE $M_i \leftarrow $ consult $i$-th line of table
		\IF {$a \ge M+M_i$}
		\STATE $M\leftarrow M+M_i$
		\STATE $i\;\;\,\leftarrow i+1$
		\ELSE
		\STATE $\eta\;\leftarrow\frac{\pi}{4}+2i\arcsin\frac{d}{2}$
		\STATE $m\leftarrow$ as in Proposition \ref{prop:T_eta}, item \ref{prop:m})
		\STATE $n\;\leftarrow$ as in Proposition \ref{prop:T_eta}, item \ref{prop:n})
		\STATE $j\,\:\leftarrow (a-M) \mod m $
		\STATE $k\,\:\leftarrow \lfloor (a-M)/m \rfloor$
		\STATE $\xi_1 \leftarrow j\frac{2\pi}{m} +k\frac{\pi}{m}$
		\vspace{0.5mm}
		\STATE $\xi_2 \leftarrow k\frac{2\pi}{n}$
		\RETURN $\mathbf{x} \leftarrow$\\ $(\cos\eta\cos\xi_1,\cos\eta\sin\xi_1,\sin\eta\cos\xi_2,\sin\eta\sin\xi_2)$	
		\ENDIF
		\ENDFOR
	\end{algorithmic} 
\end{algorithm}

Each individual line of Algorithm~\ref{alg-enc-4} has constant complexity (constant number of additions, multiplications, trigonometric functions etc.). In the worst-case scenario, the main loop (line~2) will be executed $t=t(d)$ times. Note that
\begin{displaymath}
t(d) = \left\lfloor\frac{\pi}{4\arcsin(d/2)}\right\rfloor \le \frac{\pi}{4\arcsin(d/2)} \le \frac{\pi}{2d},
\end{displaymath}
so the computational complexity of the algorithm is $O(t) = O(d^{-1})$.

\subsection{General Case: Spherical Codes in $\mathbb{R}^{2n}$}

We consider now the algorithm that implements the map $F(a)=(x_1,\dots,x_{2n})\in\mathcal{C}(M,2n,d)$. This injection can be decomposed with the help of the following maps:
\begin{align}	\label{eq:sigma}
	\begin{split}
	\sigma : \Sigma &\to \left[0,\frac{\pi}{2}\right] \times \Sigma_1 \times \Sigma_2\\
	a &\mapsto (\eta; a_1, a_2),
	\end{split}
\end{align}

\begin{align} \label{eq:iotachi}
	\begin{split}
	\iota\circ\chi : \Sigma &\to \mathcal{C}(M,4,d)\\
	a &\mapsto (\cos\eta\cos\xi_1,\cos\eta\sin\xi_1,\sin\eta\cos\xi_2,\sin\eta\sin\xi_2),
	\end{split}	
\end{align}
and
\begin{align} \label{eq:Phi}
	\begin{split}
	\hspace{5em}&\hspace{-5em}\Phi : \left[0,\frac{\pi}{2}\right] \times \mathcal{C}(|\Sigma_1|,n,d/\cos\eta) \times \mathcal{C}(|\Sigma_2|,n,d/\sin\eta)\\
	&\to \mathcal{C}(|\Sigma_1||\Sigma_2|,2n,d)\\
	\hspace{5em}&\hspace{-5em}\left(\eta;\,(x_1,\dots,x_n),\,(y_1,\dots,y_n)\right)\\
	&\mapsto \left(\cos\eta\,(x_1,\dots,x_n);\,\sin\eta\,(y_1,\dots,y_n)\right),
	\end{split}
\end{align}
where $\Sigma$ is the alphabet of the total code ($|\Sigma|=M$) and $\Sigma_1$, $\Sigma_2$ are the alphabets of each half-dimension code.

\begin{figure}[!t]
	\centering
	\includegraphics[width=7cm]{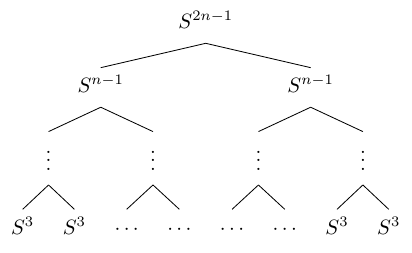}
	\caption{Decomposition tree for $S^{2n-1}$. Note that the number of nodes is $\sum_{i=0}^{k-2} 	2^i = 2^{k-1}-1 = n-1$, with $k=\log_2(2n)$.}
	\label{fig:decomposition-tree}
\end{figure}

We also assume that we have stored tables with information on each leaf $\left(S^{\tilde{n}-1}\times S^{\tilde{n}-1}\right)_\eta$, for dimensions $\tilde{n} \in \{n, n/2, \dots, 2\}$, during the construction of the code. Each line of such a table contains the index $i$ of the leaf, its parameter $\eta_i$, and the number of points $M_{i,1}$ and $M_{i,2}$ on each of the half-dimension spheres $S^{\tilde{n}-1}_{\cos\eta_i}$ and $S^{\tilde{n}-1}_{\sin\eta_i}$, respectively (see Table~\ref{tab: ex-encoding}).

\begin{table}[!h]
	\renewcommand{\arraystretch}{1.3}
	\centering
	\caption{Example of Storage Table for $\mathcal{C}(M,2n,d)$}
	\label{tab: ex-encoding}
	\begin{tabular}{cccc}
		\hline
		$i$      & $\eta_i$             & $M_{i,1}$  & $M_{i,2}$  \\ \hline \hline
		$\vdots$ & $\vdots$           & $\vdots$   & $\vdots$   \\ 
		$-1$     & $\pi/4-\Delta\eta$ & $M_{-1,1}$ & $M_{-1,2}$ \\ 
		$0$      & $\pi/4$            & $M_{0,1}$  & $M_{0,2}$  \\ 
		$1$      & $\pi/4+\Delta\eta$ & $M_{1,1}$  & $M_{1,2}$  \\ 
		$\vdots$ & $\vdots$           & $\vdots$   & $\vdots$   \\ \hline
	\end{tabular}
\end{table}

The length of each  table is equal to the number $\tilde{t}$ of leaves in the corresponding dimension. For $S^{2n-1}$, this number is $t(d) = \lfloor \pi/4\arcsin (d/2) \rfloor$ (Corollary~\ref{cor: t}). There will be $n-1$ tables, one for each sphere (node) of the decomposition tree (Fig. \ref{fig:decomposition-tree}). Note that, when halving the dimension of the sphere ($S^{2n-1} \to S^{n-1}$), the size of the table in the new dimension cannot increase:
\begin{displaymath}
	d \le \min\left\{\frac{d}{\cos\eta}, \frac{d}{\sin\eta}\right\},
\end{displaymath}
therefore
\begin{align*}
	\hspace{3em}&\hspace{-3em} \left\lfloor\frac{\pi}{4\arcsin(d/2)}\right\rfloor \ge\\
	&\max\left\{ \left\lfloor\frac{\pi}{4\arcsin(d/2\cos\eta)}\right\rfloor, \left\lfloor\frac{\pi}{4\arcsin(d/2\sin\eta)}\right\rfloor \right\}.
\end{align*}
Thus the storage space needed is no greater than $(n-1)t(d)$, and the storage complexity is $O(nt) = O(nd^{-1})$, which is linear in the dimension $n$.

\begin{algorithm}[!t]
	\caption{Algorithm implementing map $\sigma$.}
	\label{alg-sigma-2n}
	\begin{algorithmic}[1]
		\renewcommand{\algorithmicrequire}{\textbf{Input:}}
		\renewcommand{\algorithmicensure}{\textbf{Output:}}
		\REQUIRE $a$, $d$
		\ENSURE $\eta$, $a_1$, $a_2$
		
		\STATE $M \leftarrow$ 0
		\STATE $\tilde{t} \leftarrow \left\lfloor \pi/4\arcsin(d/2)\right\rfloor$
		\FOR {$i\in\llbracket -\lfloor \tilde{t}/2 \rfloor,\lfloor \tilde{t}/2 \rfloor \rrbracket$}
		\STATE $(M_{i,1}, M_{i,2}) \leftarrow $ consult $i$-th line of table
		\IF{$a>M +M_{i,1}M_{i,2}$}
		\STATE $M\leftarrow M+M_{i,1}M_{i,2}$
		\STATE $i\;\;\,\leftarrow i+1$
		\ELSE
		\STATE $\eta\:\,\leftarrow\frac{\pi}{4}+2i\arcsin\frac{d}{2}$
		\STATE $a_1 \leftarrow (a-M)\mod M_{i,1}$
		\STATE $a_2 \leftarrow \lfloor (a-M)/M_{i,1}\rfloor$
		\RETURN {$(\eta;a_1,a_2)$}
		\ENDIF
		\ENDFOR
	\end{algorithmic} 
\end{algorithm}

\begin{algorithm}[!t]
	\caption{Encoding algorithm for $\mathbb{R}^{n}$ (map $F$).}
	\label{alg-F-2n}
	\begin{algorithmic}[1]
		\renewcommand{\algorithmicrequire}{\textbf{Input:}}
		\renewcommand{\algorithmicensure}{\textbf{Output:}}
		\REQUIRE $n$, $a$, $d$
		\ENSURE $\mathbf{x}=(x_1,\dots,x_n)$
		\IF {$n=4$}
		\RETURN $\mathbf{x}\leftarrow \iota\circ\chi(a,d)$
		\ELSE
		\STATE $(\eta;a_1,a_2)\leftarrow\sigma(a,d)$
		\STATE $(w_1,\dots,w_{n/2}) \leftarrow F(n/2, a_1, d/\cos\eta)$
		\STATE $(z_1,\dots,z_{n/2}) \;\;\leftarrow F(n/2, a_2, d/\sin\eta)$
		\RETURN $\mathbf{x} \leftarrow \left(\cos\eta\, (w_1,\dots,w_{n/2}),\, \sin\eta\, (z_1,\dots,z_{n/2})\right)$
		\ENDIF
	\end{algorithmic} 
\end{algorithm}

The algorithm that implements the map $\iota\circ\chi$ in \eqref{eq:iotachi} is the encoding algorithm for the basic case $\mathbb{R}^4$ (Algorithm~\ref{alg-enc-4}), and it has complexity $O(t)$. The map $\sigma$ in \eqref{eq:sigma} is implemented by Algorithm~\ref{alg-sigma-2n}. Each individual line has constant complexity and, in the worst-case scenario, the main loop (line~3) is repeated $\tilde{t} \le t$ times, hence it has complexity $O(t) = O(d^{-1})$.

Finally, the implementation of map $F$ is represented in Algorithm~\ref{alg-F-2n}. The general step for dimension $n$ computes $\sigma(a,d)$ with $O(d^{-1})$ (line~4), calls itself twice with parameter $n/2$ (lines~5 and~6), and performs $n$ multiplications (line~7). If we have a good family of codes in dimension $n/2$, with encoding complexity $O(f(n))$, and we apply one iteration of Algorithm \ref{alg-F-2n} to double the dimension with SCHF construction, the encoding complexity of the new code with respect to the dimension $n$ will be $O(\max\{2f(n), n\})$. In the recursive case, the number of steps of the recurrence is characterized by
\begin{displaymath}
T(n) = 2 T\left( \frac{n}{2} \right) + O(n) + O(d^{-1}).
\end{displaymath}
Using the master theorem \cite[p. 73]{cormen}, we find that, for fixed $d$, this algorithm has complexity $O(n \log n)$.

We can compare this complexity with known TLSC implementations~\cite{naves}. Codes obtained via a subcode with $k$ elements have linear time complexity; in spite of the low complexity, they have the weakest performance among TLSC implementations and are outperformed by recursive SCHF in most scenarios (see Section~\ref{sec:performance}). Codes on polygon layers have the best performance among TLSC implementations and the closest to SCHF; nonetheless, their exact complexity has not been established and, based on the code structure and computing time required for tested examples, seems to be higher than recursive SCHF. For instance, the results in Table~\ref{tab: TLSC-higher} for this construction, with $d=0.01$ in dimensions 16 and 32, could not be computed using the implementation provided in~\cite{naves} under the same time and storage resources as the other two codes.

\section{Decoding} \label{sec:decoding}

Given a vector $\mathbf{y} \in \mathbb{R}^{n}$ and a spherical code $\mathcal{C}(M,n,d)$, the maximum likelihood~(ML) decoding consists in finding the vector $\mathbf{x}\in\mathcal{C}(M,n,d)$ such that
\begin{equation}
\mathbf{x} = \arg\min_{\mathbf{x}_i\in\mathcal{C}}\|\mathbf{y}-\mathbf{x}_i\|.
\end{equation}
As shown in~\cite{torezzan}, to decode a received vector $\mathbf{y}$ in a spherical code, we can consider $\mathbf{y}$ to be a unit vector and the problem is equivalent to
\begin{equation}
\label{decode}
\mathbf{x} = \arg\max_{\mathbf{x}_i\in\mathcal{C}}\langle\mathbf{x}_i,\mathbf{y}\rangle.
\end{equation}
For small codes, it is feasible to obtain \eqref{decode} by computing all inner products and choosing the maximizing codeword. But, to avoid high-complexity of ML decoding on larger codes, we introduce a sub-optimal decoding algorithm for standard SCHF construction, which is inspired by~\cite{torezzan} and does not require storage of the whole codebook. As previously, we start with a procedure for the basic case $\mathbb{R}^4$ and then generalize it recursively to $\mathbb{R}^{2n}$.

\subsection{Basic Case: Spherical Codes in $\mathbb{R}^4$}

Using the Hopf foliation, a unit vector $\mathbf{y}=(y_1,y_2,y_3,y_4)$ may be written as
\begin{displaymath}
	(y_1+\boldsymbol{i}y_2,\, y_3+\boldsymbol{i}y_4) = (e^{\boldsymbol{i}\xi_1}\cos\eta,\, e^{\boldsymbol{i}\xi_2}\sin\eta),
\end{displaymath}
where
\begin{align}
	\eta &= \arctan\left(\sqrt{\frac{y_3^2+y_4^2}{y_1^2+y_2^2}}\right), \label{eq: decode1}\\
	\xi_1 & =\arctan(y_2/y_1), \label{eq: decode2}\\
	\xi_2 &= \arctan(y_4/y_3) \label{eq: decode3}.
\end{align}
In general, however, the triplet  $(\eta;\xi_1,\xi_2)$ does not parametrize a point of the codebook. So our objective is to find the triplet $(\hat{\eta};\hat{\xi_1},\hat{\xi_2})$ which parametrizes the codeword closest to~$\mathbf{y}$. Let us denote our guess by $\hat{\mathbf{x}}=(e^{\boldsymbol{i}\hat\xi_1}\cos\hat\eta,\,e^{\boldsymbol{i}\hat\xi_2}\sin\hat\eta)$. We propose a two-step decoding method, as follows.

\begin{enumerate}
	\item The first step is to search for the torus $T_{\hat{\eta}}$ closest to received point~$\mathbf{y}$. Thanks to Proposition~\ref{prop:dist-tori}, this is equivalent to finding
	\begin{equation} \label{eq:closest-eta}
	\hat{\eta} = \arg\min_{\eta' \in H} |\eta' - \eta |,
	\end{equation}
	where $H = \{ \frac{\pi}{4} + 2i\arcsin\frac{d}{2},\ -\lfloor t(d)/2 \rfloor \le i \le \lfloor t(d)/2 \rfloor \}$ is the set of $\eta$-parameters used in the code.
	
	\item Once $\hat{\eta}$ is determined, we project $\mathbf{y}$ on $T_{\hat{\eta}}$, obtaining $(e^{\boldsymbol{i}\xi_1}\cos\hat{\eta},\,e^{\boldsymbol{i}\xi_2}\sin\hat{\eta})$. This is the point on $T_{\hat{\eta}}$ which is closest to the received vector $\mathbf{y}$. To obtain $\hat{\xi}_1$ and $\hat{\xi}_2$, we compute
	\begin{equation}
	\hat{k} = \lfloor \xi_2/\Delta\xi_2 \rceil \mod n \label{eq-k-hat}
	\end{equation}
	and
	\begin{equation}
	\hat{j} = \left\lfloor \frac{ \xi_1 - \hat{k}\Delta\xi_1/2 }{\Delta\xi_1} \right\rceil \mod m, \label{eq-j-hat}
	\end{equation}
	where $\lfloor \cdot \rceil$ denotes the rounding function and $\Delta\xi_1=2\pi/m$, $\Delta\xi_2=2\pi/n$. Then,
	\begin{align}
	\hat{\xi}_1 &= \hat{j}\Delta\xi_1 + \hat{k}\Delta\xi_2 \label{eq-xi1-hat},\\
	\hat{\xi}_2 &= \hat{k}\Delta\xi_2 \label{eq-xi2-hat}.
	\end{align}
\end{enumerate}

\begin{algorithm}[!t]
	\caption{Decoding algorithm in $\mathbb{R}^4$.}
	\label{alg: decodeR4}
	\begin{algorithmic}[1]
		\renewcommand{\algorithmicrequire}{\textbf{Input:}}
		\renewcommand{\algorithmicensure}{\textbf{Output:}}
		\REQUIRE $\mathbf{y}$, $d$
		\ENSURE $\hat{\mathbf{x}} = (\hat{x}_1, \hat{x}_2, \hat{x}_3, \hat{x}_4)$
		
		\STATE $\mathbf{y} \leftarrow \mathbf{y}/\|\mathbf{y}\|$
		\STATE $(\eta,\xi_1,\xi_2) \leftarrow$ as in \eqref{eq: decode1}, \eqref{eq: decode2}, \eqref{eq: decode3}
		\STATE $\hat{\eta}\; \leftarrow \left\lfloor\frac{\eta-\pi/4}{\Delta\eta}\right\rceil \Delta\eta + \frac{\pi}{4}$, with $\Delta\eta=2\arcsin(d/2)$
		\STATE $m\leftarrow$ as in Proposition \ref{prop:T_eta}, item \ref{prop:m})
		\STATE $n\;\leftarrow$ as in Proposition \ref{prop:T_eta}, item \ref{prop:n})
		\STATE $(\hat\xi_1, \hat\xi_2) \leftarrow$ as in \eqref{eq-k-hat}, \eqref{eq-j-hat}, \eqref{eq-xi1-hat}, \eqref{eq-xi2-hat}
		\RETURN $\hat{\mathbf{x}} \leftarrow$\\ $(\cos\hat\eta\cos\hat\xi_1,\cos\hat\eta\sin\hat\xi_1,\sin\hat\eta\cos\hat\xi_2,\sin\hat\eta\sin\hat\xi_2)$
	\end{algorithmic} 
\end{algorithm}

These steps are detailed in Algorithm~\ref{alg: decodeR4}. To further approach the minimum distance solution, additional steps can be considered. If $\hat{d} \coloneqq \|\hat{\mathbf{x}}-\mathbf{y}\|<d/2$, the decoding is finished. Otherwise, the closest point $\mathbf{x}^*$ may be on another torus $T_{\eta*}$. We can look for the set of tori with parameters $\{\eta_1,\dots,\eta_\nu\}$ for which $d_i=\|\hat{\mathbf{x}}_i-\mathbf{y}\|<\hat{d}$, where
\begin{displaymath}
	\hat{\mathbf{x}}_i = \left(e^{\boldsymbol{i}\xi_{1,i}}\cos\eta_i,\,e^{\boldsymbol{i}\xi_{2,i}}\sin\eta_i\right)
\end{displaymath}
is obtained from the projection on torus $T_{\eta_i}$, $i\in\{1,\dots,\nu\}$ and we choose $\mathbf{x}^*=\hat{\mathbf{x}}_i$ that minimizes $d_i$. As for the coding design, in dimension 4, this procedure approaches the one proposed for TLSC~\cite{torezzan}.

\subsection{General Case: Spherical Codes in $\mathbb{R}^{2n}$}

Let us generalize the decoding procedure to codes on $S^{2n-1}$. As before, write the arbitrary received vector $\mathbf{y}\in S^{2n-1}$ as
\begin{displaymath}
\mathbf{y}=\left(\cos\eta\,(y_1,\dots,y_n),\,\sin\eta\,(y_{n+1},\dots,y_{2n})\right),
\end{displaymath}
with
\begin{equation} \label{eq: decode4}
\eta=\arctan\left(\sqrt{\frac{\sum_{i=n+1}^{2n}y_i^2}{\sum_{j=1}^{n}y_j^2}}\right).
\end{equation}

A similar two-step procedure can be deduced, as follows.

\begin{enumerate}
	\item Find the closest leaf $\left(S^{n-1}\times S^{n-1}\right)_{\hat{\eta}}$ to point~$\mathbf{y}$, i.e., the value~$\hat{\eta}$ that parametrizes a leaf used in the code closest to~$\eta$, as in \eqref{eq:closest-eta}.
	
	\item We can then split $S^{2n-1}$ into $S_{\cos\hat{\eta}}^{n-1}$ and $S_{\sin\hat{\eta}}^{n-1}$ and recursively apply the procedure all the way down to the basic case $S^3\subset\mathbb{R}^4$. The additional steps can also be applied for each leaf $\left(S^{n-1}\times S^{n-1}\right)_{\eta}$.
\end{enumerate}

A pseudocode for the recursive method is presented in Algorithm~\ref{alg-dec-geral}.

\begin{algorithm}[!t]
	\caption{Decoding algorithm in $\mathbb{R}^{n}$.}
	\label{alg-dec-geral}
	\begin{algorithmic}[1]
		\renewcommand{\algorithmicrequire}{\textbf{Input:}}
		\renewcommand{\algorithmicensure}{\textbf{Output:}}
		\REQUIRE $\mathbf{y}$, $d$, $n$
		\ENSURE $\hat{\mathbf{x}} = (\hat{x}_1, \dots, \hat{x}_n)$
		\IF {$n=4$}
		\STATE Apply Algorithm~\ref{alg: decodeR4} to $\mathbf{y}$ and $d$
		\ELSE
		\STATE $\mathbf{y} \leftarrow \mathbf{y}/\|\mathbf{y}\|$
		\STATE Compute $\eta$ from $\mathbf{y}$ using~\eqref{eq: decode4}
		\STATE $\hat{\eta} \leftarrow \left\lfloor\frac{\eta - \pi/4}{\Delta\eta}\right\rceil \Delta\eta + \frac{\pi}{4}$, with $\Delta\eta=2\arcsin(d/2)$
		\STATE $(w_1,\dots,w_{n/2}) \leftarrow$ apply decoding in $\mathbb{R}^{n/2}$ to $(y_1,\dots,y_{n/2})$ with $d/\cos\hat\eta$
		\STATE $(z_1,\dots,z_{n/2}) \:\leftarrow$ apply decoding in $\mathbb{R}^{n/2}$ to $(y_{(n/2) + 1},\dots,y_{n})$ with $d/\sin\hat\eta$
		\RETURN{$\hat{\mathbf{x}} \leftarrow \left(\cos\hat\eta\, (w_1,\dots,w_{n/2}),\, \sin\hat\eta\, (z_1,\dots,z_{n/2})\right)$}
		\ENDIF
	\end{algorithmic} 
\end{algorithm}

\subsection{Decoding Performance}

We analyze the performance of decoding using the previously presented standard SCHF procedure. The only information required to store are minimum distance and dimension, so this algorithm has storage complexity $O(1)$.

The number of operations $T(n)$ in the general loop of Algorithm~\ref{alg-dec-geral} follows the recursive expression
\begin{displaymath}
T(n) = 2 T\left( \frac{n}{2} \right) + O(n),
\end{displaymath}
where $O(n)$ accounts for computing $\mathbf{y}$ (line~4), $\eta$ from~\eqref{eq: decode4} (line~5) and the products in line~9. Using the master theorem~\cite[p. 73]{cormen}, it follows that this algorithm has complexity $O(n \log n)$. Compare this with the complexity of a brute-force ML decoder, which has time complexity $O(Mn)$ and storage complexity $O(M)$. In~\cite{hamkins3}, two decoding algorithms are proposed for laminated spherical codes: one uses $O(\sqrt{M})$ space and $O(\log M)$ time and the other uses $O(1)$ space and $O(\sqrt{M})$ time.

\begin{figure*}[!t]
	\centering
	\subfloat[$\mathcal{C}(52,4,0.7)$]{\includegraphics[width=5cm]{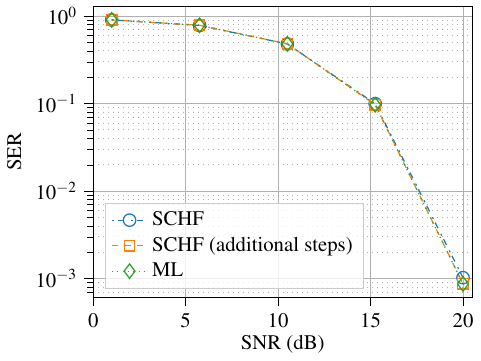}}
	\hfil
	\subfloat[$\mathcal{C}(152,4,0.5)$]{\includegraphics[width=5cm]{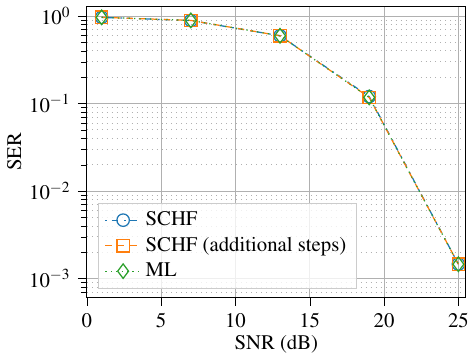}}
	\hfil
	\subfloat[$\mathcal{C}(360,8,0.7)$]{\includegraphics[width=5cm]{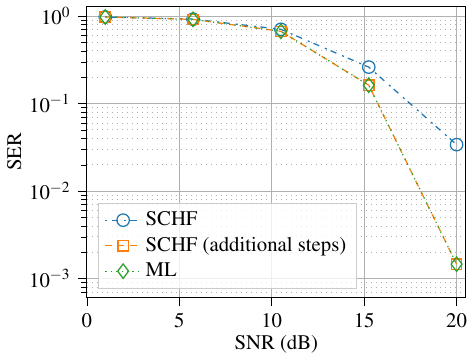}}
	\caption{Symbol error rate~(SER) for decoding different SCHF $\mathcal{C}(M,n,d)$ with sub-optimal SCHF methods and brute-force ML decoder.}
	\label{fig: decoding-rates}
\end{figure*}

To analyze the performance of this sub-optimal decoder, we have performed the following test: for a given code $\mathcal{C}(M,n,d)$, we add i.i.d. centered Gaussian noise $\mathbf{z}_i \sim \mathcal{N}(0,\sigma^2)$ to each point~$\mathbf{x}_i$ in the code and decode each $\mathbf{y_i} = \mathbf{x}_i + \mathbf{z}_i $. We compute the symbol error rate~(SER) for different signal-to-noise ratios~(SNR), as well as the average CPU time\footnote{
	Using \textit{Python 3.7.6} on a 8GB RAM, Intel Core i5-7200U @ 2.50GHz machine.
} required for decoding one codeword using the proposed algorithm without additional steps, with some additional steps (if $\hat{d} \ge d/2$, we consider the two adjacent leaves $\eta \pm \Delta\eta$), and the brute-force ML decoder. Results are presented in Fig.~\ref{fig: decoding-rates} and Table~\ref{tab: decoding-times}.

While the SER of the proposed decoder without additional steps is higher than brute-force ML for higher SNR, the average time required to decode one codeword by the latter method can be up to ten times higher. On the other hand, when allowing simple additional steps, the decoding performance practically matches brute-force ML, while keeping low time complexity. This justifies the use of the proposed sub-optimal decoder.

\begin{table}[!t]
	\renewcommand{\arraystretch}{1.3}
	\centering
	\caption{Average CPU Time ({\upshape ms}) for Decoding One Codeword}
	\label{tab: decoding-times}
	\centering
	\begin{tabular}{cccc}
		\hline
		& \makecell{SCHF without\\ additional steps} & \makecell{SCHF with\\ additional steps} & Brute-force ML \\ \hline \hline
		$\mathcal{C} (52, 4, 0.7)$  & $0.109$ & $0.139$ & $0.409$ 		 \\
		$\mathcal{C} (152, 4, 0.5)$ & $0.114$ & $0.139$ & $1.169$        \\
		$\mathcal{C} (360, 8, 0.7)$ & $0.287$ & $0.582$ & $2.837$        \\ \hline
	\end{tabular}
\end{table}

\section{Conclusion} \label{sec:conclusion}

We propose a construction for spherical codes in dimensions~$2^k$ by a recursive procedure that is based on the Hopf foliations of $S^{2n-1}$ by $\left(S^{n-1}\times S^{n-1}\right)_{\eta}$ and uses $\mathbb{R}^4$ as basic case. In fact, this construction can be applied to any even dimension~$2n$ as long as a family of spherical codes is provided in dimension~$n$.

Given a minimum distance $d>0$, the standard method chooses leaves $S_{\cos \eta}^{n-1} \times S_{\sin\eta}^{n-1}$, parametrized by $\eta \in \left[0, \pi/2\right]$, that foliate $S^{2n-1}$ while mutually distant by at least~$d$. On each leaf, we recursively distribute points on each of the spheres $S_{\cos\eta}^{n-1}$ and $S_{\sin\eta}^{n-1}$, with scaled minimum distances and combine the results as a Cartesian product. In the basic case $\mathbb{R}^4$, the sphere~$S^3$ is foliated by tori~$T_\eta$, each of which is divided in internal circles mutually distant by~$d$, where points are equidistantly distributed.

In non-asymptotic regime, SCHF compare favorably to other constructive methods. Asymptotic upper bounds for the recursive and half-dimension SCHF are derived and compared with  other constructions. An encoding algorithm is presented, the time and storage complexities of which are respectively $O(n \log n)$ and  $O(n)$. A sub-optimal decoder with time complexity $O(n \log n)$ and storage complexity $O(1)$ is also proposed. We verify in some examples that, by allowing additional steps, its SER is close to that of ML decoder, while keeping the time required significantly lower.

Perspectives for the extension of this work include investigating, in several $2n$ dimensions, SCHF constructed from good available codes in dimension~$n$; considering the structure of quaternions and octonions in the construction of codes; and analyzing the proposed SCHF for vector quantitation of Gaussian sources.

\section*{Acknowledgment}

The authors are grateful for some generous contributions by C.~Torezzan, V.~Vaishampayan and L.~Naves, as well as to J.~Hamkins and K.~Zeger, for sharing their apple-peeling implementation. We also thank the referees for their important suggestions, which have meaningfully improved the original manuscript.

\ifCLASSOPTIONcaptionsoff
  \newpage
\fi




\end{document}